\documentclass[%
aip,
 amsmath,amssymb,
reprint, onecolumn 
]{revtex4-2}

\draft 

\usepackage{amsmath}
\usepackage{amsfonts}
\usepackage{amssymb}
\usepackage{color}
\usepackage{hyperref}
\usepackage{graphicx}
\usepackage{dcolumn}
\usepackage{bm}

\allowdisplaybreaks
\hypersetup{
    colorlinks=true,
    linkcolor=magenta,
    filecolor=blue,      
    citecolor=blue
}

\makeatletter
\def\@email#1#2{%
 \endgroup
 \patchcmd{\titleblock@produce}
  {\frontmatter@RRAPformat}
  {\frontmatter@RRAPformat{\produce@RRAP{*#1\href{mailto:#2}{#2}}}\frontmatter@RRAPformat}
  {}{}
}%
\makeatother

\begin{document}


\title
{STEGR in Internal-Space Formulation: Formalisms, Primary Constraints, and Possible Internal Symmetries}
\author{Kyosuke TOMONARI}
\email{ktomonari.phys@gmail.com}
\affiliation{Department of Physics,\\
Institute of Science Tokyo,\\
2-12-1 Ookayama, Meguro-ku, Tokyo 152-8551, Japan}
\affiliation{Interfaculty Initiative in Information Studies, Graduate School of Interdisciplinary Information Studies, The University of Tokyo,\\
Hongo Campus, 7-3-1 Hongo, Bunkyo-ku, Tokyo, Japan\\}

\date{\today}
             
\begin{abstract}
We establish the theories of Symmetric Teleparallel Equivalent to General Relativity (STEGR) in the internal-space and investigate possible internal-space symmetries among primary constraint densities in the theories. First of all, we revisit STEGR in terms of the gauge approach to gravity and formulate it in the internal-space set-up. We find three possible formalisms according to the vanishing-torsion property. Then, we investigate possible internal-space symmetries in each formalism. We find that in our formulation there are two possible symmetries. One satisfies the translation symmetry but broken in the local symmetry provided by the general linear group which contains the local Lorentz symmetry. The other satisfies the latter symmetry but is absent in the former symmetry. Finally, we conclude this work and show future perspectives. 
\end{abstract}

\maketitle




\section{\label{01}Introduction}
General Relativity (GR) is the most successful theory to describe the wide range of gravitational phenomena in terms of pseudo-Riemannian geometry based on the local Lorentz invariance, the diffeomorphism symmetry, and Einstein's equivalence principle. From the physical point of view, however, there is no reason to restrict our theories to this particular geometry. In fact, Einstein himself reconstructed GR in an alternative way using another geometry based purely on torsion instead of curvature, labeled as teleparallel gravity~\cite{Einstein1928}. For a detailed review on teleparallel gravity, see Ref.~\cite{Bahamonde:2021gfp} and the Refs. therein. In modern perspectives, it is known that GR has its equivalent formulation of the so-called geometrical Trinity of Gravity (ToG), in which gravitation is treated with the torsion (Teleparallel Equivalent to GR: TEGR) and/or the non-metricity (Symmetric Teleparallel Equivalent to GR: STEGR) instead of the curvature up to boundary terms~\cite{Nester:1998mp,BeltranJimenez:2019esp,Heisenberg:2018vsk}. These two theories assume that the general curvature is vanishing. In more generic perspectives, ToG is a set of specific classes in the so-called Metric-Affine gauge theories of Gravity (MAG), which is disciplined by the gauge invariant characteristics~\cite{Utiyama:1956sy,Kibble:1961ba,Ivanenko:1983fts}. For a detailed review on MAG, see Refs.~\cite{Hehl:1994ue} and the Refs. therein. 

Modern cosmology is established based on GR and the standard model of elementary particles~\cite{Dodelson:2003ft,Mukhanov:2005sc,Weinberg:2008zzc}. Observations have been unveiled the new perspectives in the modern cosmology such as the necessity of inflation~\cite{Planck:2018vyg,Tsujikawa:2003jp,Vazquez:2018qdg}, the existence of dark matter~\cite{Freese:2008cz,Billard:2021uyg,Planck:2018vyg}, the late-time acceleration of the universe (or the existence of dark energy)~\cite{Planck:2018vyg,SupernovaSearchTeam:1998fmf,SupernovaCosmologyProject:1998vns}, and most recently the tension in the Hubble constant~\cite{Planck:2018vyg,H0LiCOW:2019pvv,Riess:2019cxk,Schoneberg:2022ggi,ACT:2023kun}. 
These issues cannot be explained basing on GR and the standard model of elementary particles, which provide the fundamental theories establishing the modern cosmology.
One of the approaches to challenge such issues is to reconsider the fundamental theory of gravity, {\it i.e.,} GR, on the ground of MAG frameworks. In particular, the non-linear extension of MAG theories in the same manner as $f(R)$-gravity is remarkable for approaching these issues. For a detailed review on the extended theories, see Refs.~\cite{Nojiri:2006ri,DeFelice:2010aj,Cai:2015emx,Heisenberg:2023lru,Zhao:2021zab} and Refs. therein. The extended theories provide the well-behaved inflation models~\cite{Buchdahl:1970ldb,Rezazadeh:2017edd,Capozziello:2022tvv,Nojiri:2024zab,Nojiri:2024hau,Gamonal:2020itt,Capozziello:2024lsz} in the high precision to the recent observations given by Planck 18~\cite{Planck:2018nkj}. Furthermore, the extended theories {\it a priori} contain an effective cosmological constant and an effective gravitational constant in their field equations, and these effective constants can explain the late-time acceleration of the universe~\cite{Tsujikawa:2010sc,Bamba:2010wb,Zubair:2015opa,Bahamonde:2017ize,Solanki:2022ccf,Nojiri:2024hau,Nojiri:2024zab} and reconcile the Hubble-tension~\cite{DiValentino:2021izs,Heisenberg:2022gqk}, respectively. These significant characteristics are ascribed to the inherent extra Degrees of Freedom (DoF) in the theories. To clarify the novel DoF of the theories, the Dirac-Bergmann analysis can be applied~\cite{Dirac:1950pj,Dirac:1958sq,Bergmann:1949zz,BergmannBrunings1949,Bergmann1950,Anderson:1951ta}.

Scrutinizing the structure formation of the universe is also one of the significant issues in the modern cosmology~\cite{Dodelson:2003ft,Mukhanov:2005sc,Weinberg:2008zzc,Matsubara:2022ohx,Matsubara:2022eui,Matsubara:2023avg,Matsubara:2024sqn}. To approach this issue, we employ the linear perturbation theory around the flat and non-flat Friedmann–Lemaître–Robertson–Walker (FLRW) spacetime~\cite{Bernardeau:2001qr,Malik:2008im}. However, in the extension/modification of the theories of gravity, there generically occurs a discrepancy in the number of the propagating DoF in the linear perturbation and that of the full DoF of a given theory. The discrepancy is called the `{\it strong coupling(s)}' around the background spacetime that is chosen for the perturbation in advance. For a detailed description, see Sec. IV in Ref.~\cite{Bahamonde:2024zkb}. Perturbation theories that suffer from this problem would not predict physical phenomena in a healthy manner due to the lack of the DoF existing in the origin theory. The extension/modification of MAG also encounters this problem. For example, see Refs.~\cite{Bahamonde:2020lsm,Bahamonde:2022ohm,Aoki:2023sum,Tomonari:2023wcs,Bahamonde:2024zkb}. To investigate whether or not the problem exists in a given theory, we have to clarify the full DoF of the theory. Again, the Dirac-Bergmann analysis plays a crucial role in this subject. 

In recent years, STEGR and its extended theories have been gathering great attention and investigating eagerly~\cite{Capozziello:2024vix,Nojiri:2024zab,Nashed:2024jqw,Nojiri:2024hau}. In particular, STEGR in the coincident gauge so-called Coincident General Relativity (CGR) is the most enthusiastically scrutinized and understood~\cite{BeltranJimenez:2017tkd,Heisenberg:2023lru}, in which all covariant derivatives are formally alternated by ordinary partial derivatives due to vanishing the affine connection. From the viewpoint of { \it healthy} application to cosmology and astrophysics, revealing the full DoF of the theories is mandatory. Here, the jargon ``{\it healthy}'' means that the theory is free not only from strong couplings but also from ghost modes around a background spacetime in a perturbation theory. However, there is no complete Dirac-Bergmann analysis not only in STEGR but also in its extensions including that of the non-linear extension such as $f(Q)$-gravity, even not in those in the cases of CGR. For instance, the Dirac-Bergmann analysis on the coincident $f(Q)$-gravity is still an open problem. In Ref.~\cite{Hu:2022anq}, on one hand, it was revealed that the full DoF is calculated in eight. The authors in Ref.~\cite{DAmbrosio:2023asf} addressed that the consistency conditions, which determine the Lagrange multipliers and whether secondary or higher-order constraint densities arise or not in the theory, may take the form not only in algebraic equations but also in Partial Differential Equations (PDEs). In particular, in the latter case, there would give rise to a case that not all multipliers are determined due to the problem of the solvability of the PDEs, not to the existence of first-class constraint densities. To prevent this problem, in Ref.~\cite{Tomonari:2023wcs}, a `{\it prescription}' was proposed, in which the terms to make the consistency conditions to be PDEs are removed, and the method allows us to analyze a sector which is determined by solving the conditions only in the form of algebraic equations. Then, it was unveiled that the theory bifurcates in several sectors, and the DoF is calculated as six in a generic sector under the prescription and also possible to be taken as seven without the prescription and, five and null under the prescription in special sectors. Such bifurcations occur under the broken symmetry~\cite{Blagojevic:2020dyq,Blagojevic:2023fys,Tomonari:2023wcs}. In fact, in CGR the diffeomorphism symmetry is lost. On the other hand, from the perspectives of the perturbation theory, the authors in Ref~\cite{Gomes:2023tur} clarified a `{\it pathology}' that the propagating DoF is seven with one ghost DoF in the non-trivial branch I in their jargon~\cite{Gomes:2023hyk}. Namely, the prescription might not remedy the pathology. In the current paper, to change some points of view and give new insights into this sort of problem, we establish a new formulation of STEGR in terms of the internal-space set-up, in which we do not assume the imposition of the coincident gauge. We would expect that the DoFs of the theories of STEGR are verified from different viewpoints while clarifying possible constraint structures in STEGR.

From a more generic perspective, in order to obtain the teleparallel theories of gravity from the generic metric-affine gauge theories of gravity, we must realize the teleparallel condition which demands vanishing the generic curvature as a gauge condition. The Weitzenb\"{o}ck gauge condition is such a condition to realize it as a sufficient condition, but in this gauge, the non-metricity automatically vanishes. This means that in this gauge the STEGR theories cannot be formulated in a well-posed manner as it is due to this automatic vanishment of the non-metricity. Thus, we must reformulate the theory to work well in this gauge. This paper is devoted also to resolving this issue and provides a well-posed reformulation of STEGR in this gauge.

The construction of the current paper is as follows. In Sec.~\ref{02}, we revisit STEGR in the gauge approach to gravity and establish that in the internal-space formulation. We find three possible formalisms, labeled as the internal STEGR in Formalism 1, 2, and 3. In Sec.~\ref{03}, we investigate internal symmetries in Formalism 1 and 2 by finding out primary constraint densities of which PB-algebra shows the specific algebra of the symmetries that are anticipated from the author's previous work~\cite{Tomonari:2023ars}. We reveal in $(n + 1)$-dimensional spacetime that (i) Formalism 1 can have the translation symmetry represented by $T(n+1\,:\,\mathbb{R})$ and (ii) Formalism 2 can have the local symmetry that is provided by the general linear group $G(n+1\,:\,\mathbb{R})$. Finally, in Sec.~\ref{04}, we conclude this work and give future perspectives. 

Throughout this paper, we use units with $\kappa=c^{4}/16\pi G_{N}:=1$. In the Dirac-Bergmann analysis, we denote ``$\approx$'' as the weak equality~\cite{Dirac:1950pj,Dirac:1958sq}. For quantities computed from the Levi-Civita connection, we use an over circle, ``$\,\overset{\circ}{}\,$'', on top whereas, for a general connection, tilde, ``$\,\,\tilde{}\,\,$'', is introduced. Also, Greek indices, $\alpha\,,\beta\,,\gamma\,,\cdots\,,\mu\,,\nu\,,\rho\,,\cdots\,$, denote spacetime indices whereas small Latin ones, $a\,,b\,,c\,,\cdots\,$, the spatial indices in a tangent space. Capital Latin letters, $A\,,B\,,C\,,\cdots$, are introduced to denote the internal-space indices.

\section{\label{02}Revisiting STEGR}
\subsection{\label{02:01}Gauge approach to STEGR}
Teleparallel theories of gravity is a set of special classes in more generic theory labeled as MAG~\cite{Bahamonde:2021gfp}, and MAG is formulated based on the framework of gauge approach to gravity~\cite{Hehl:1994ue}. STEGR is a more special class of teleparallel theories of gravity.

Gauge approach to gravity demands two vector bundles~\cite{Tomonari:2023wcs,Tomonari:2023ars}: a tangent bundle $(T\mathcal{M}\,,\mathcal{M}\,,\pi)$ and an internal bundle $(\mathcal{V}\,,\mathcal{M}\,,\rho)\,$, where $\mathcal{M}$ is a spacetime manifold with dimension $n+1\,$, $\pi$ is an onto map from $T\mathcal{M}$ to $\mathcal{M}\,$, and $\rho$ is an onto map from $\mathcal{V}$ to $\mathcal{M}\,$. The total space of the internal bundle, $\mathcal{V}\,$, is called merely an internal-space and, in usual formulation, it is taken to be $\mathcal{M}\times\mathbb{R}^{n+1}\,$. In this article, we obey also this ordinary choice of the internal bundle but {\it restricting only to} a local region $U\subset\mathcal{M}\,$. This means that $\left.\mathcal{V}\right|_{U} \simeq U\times\mathbb{R}^{n+1}$ but in a global region $\mathcal{V}$ is not decomposed into such simple structure. Then we introduce a frame field $\mathbf{e}\,:\,U\times\mathbb{R}^{n+1}\rightarrow T\mathcal{M}|_{U}\,$. In component form, for a basis $\zeta_{A}$ on $\left.\mathcal{M}\times\mathbb{R}^{n+1}\right|_{U} = U\times\mathbb{R}^{n+1}\,$, we can express the frame field $\mathbf{e}$ as follows: $e_{A}(p)=\mathbf{e}(p)(\zeta_{A})=e_{A}{}^{\mu}(p)\partial_{\mu}\,$, where $p\in U\,$. Note here the local property: $\left.\mathcal{M}\times\mathbb{R}^{n+1}\right|_{U} = U\times\mathbb{R}^{n+1}\simeq \left.T\mathcal{M}\right|_{U}\,$, in particular $\left.\mathcal{M}\times\mathbb{R}^{n+1}\right|_{p\in U} = \{p\}\times\mathbb{R}^{n+1} = \mathbb{R}^{n+1}\simeq\left.T\mathcal{M}\right|_{p\in U}\,$. The co-frame field of $\mathbf{e}$ is then defined as the inverse map of $\mathbf{e}$ as follows: $\mathbf{e}^{-1}\,:\,\left.T\mathcal{M}\right|_{U}\rightarrow U\times\mathbb{R}^{n+1}\,$. In component form, for the dual basis of $\zeta_{A}$, {\it i.e.,} $\zeta^{A}\,$, we have $\theta^{A}(p)=(\mathbf{e}^{-1})^{*}(p)(\zeta^{A})=\theta^{A}{}_{\mu}(p)\,dx^{\mu}\,$, where $(\mathbf{e}^{-1})^{*}$ denotes the pullback of $\mathbf{e}^{-1}$ and $p\in U\,$. Remark in general that the inverse map of the frame field can be defined only in a local region of the spacetime. The frame field and co-frame field components, $e_{A}{}^{\mu}$ and $\theta^{A}{}_{\mu}\,$, on a local region $U$ satisfies the following properties:
\begin{equation}
    e_{A}{}^{\mu}\,\theta^{A}{}_{\nu}=\delta^{\mu}{}_{\nu}\,,\quad e_{A}{}^{\mu}\theta^{B}{}_{\mu}=\delta^{A}{}_{B}\,.
\label{relations btwn e and theta}
\end{equation}
Using these ingredients, a metric $g=g_{\mu\nu}dx^{\mu}\otimes dx^{\nu}$ on $\mathcal{M}$ and a metric $\eta=\eta_{AB}\zeta^{A}\otimes\zeta^{B}$ on $\mathcal{M}\times\mathbb{R}^{n+1}$ can be related as follows: 
\begin{equation}
    g_{\mu\nu}=\theta^{A}{}_{\mu}\theta^{B}{}_{\nu}\eta_{AB}\,\quad \eta_{AB}=e_{A}{}^{\mu}e_{B}{}^{\nu}g_{\mu\nu}\,.
\label{relation between spacetime and internal space metric}
\end{equation}
Note, here, that the metric, $\eta_{AB}\,$, should be determined as a gauge in the internal-space. 

In MAG, the affine connection $\tilde{\Gamma}^{\rho}{}_{\mu\nu}$ in the spacetime $\mathcal{M}$ and the connection 1-form (components) $\omega^{A}{}_{B\mu}$ in the internal-space $\mathcal{M}\times\mathbb{R}^{n+1}$ are related in terms of 
the affine connection
as follows~\cite{Weitzenboh1923,Tomonari:2023ars}:
\begin{equation}
    \tilde{\Gamma}^{\rho}_{\mu\nu} = e_{A}{}^{\rho}\,\partial_{\mu}\theta^{A}{}_{\nu} + e_{A}{}^{\rho}\,\theta^{B}{}_{\mu}\,\omega^{A}{}_{B\nu}\,.
\label{Weitzenboch connection}
\end{equation}
Remark that this relation assumes that the local property $\left.\mathcal{M}\times\mathbb{R}^{n+1}\right|_{U}\simeq \left.T\mathcal{M}\right|_{U}$ holds. This allows us to compute the covariant derivative of the co-frame field component, $\theta^{A}{}_{\mu}\,$, as follows: $\mathcal{D}_{\nu}\theta^{A}{}_{\mu}=\partial_{\nu}\theta^{A}{}_{\mu}-\tilde{\Gamma}^{\rho}{}_{\nu\mu}\theta^{A}{}_{\rho}+\omega^{A}{}_{B\nu}\theta^{B}{}_{\mu}\,$. Applying the affine connection, Eq.~(\ref{Weitzenboch connection}), the so-called frame field postulate holds: $\mathcal{D}_{\nu}\theta^{A}{}_{\mu}=0\,$. In a generic affine connection, a Lie group action to the co-frame field provides the attribute of internal-space symmetry, or equivalently, of frame field symmetry, at each spacetime point to our theories of gravity in the usual sense of gauge theory. Namely, a co-frame field transformation $\theta^{A}{}_{\mu}\rightarrow \theta'^{A}{}_{\mu}=\Lambda^{A}{}_{B}\theta^{B}{}_{\mu}\,$, where $\Lambda^{A}{}_{B}\in G$ in which $G$ is a Lie group. In a pure geometric construction, on one hand, the internal-space symmetry given by $G$ can be taken without any restriction. In detail, see Ref.~\cite{Tomonari:2023ars}. On the other hand, in an application to physical theories, the symmetry is determined in the dependence on a given Lagrangian density. We will verify this statement throughout the current paper. Then, it leads to
\begin{equation}
    \mathcal{D}_{\mu}\theta'^{A}{}_{\nu}=\Lambda^{A}{}_{B}\mathcal{D}_{\mu}\theta^{B}{}_{\nu}
\label{gauge transformation of internal space derivative}
\end{equation}
on the ground of the transformation of the connection 1-form components as follows: $\omega^{A}{}_{B\mu}\rightarrow \omega'^{A}{}_{B\mu}=(\Lambda^{-1})^{A}_{\ C}\partial_{\mu}\Lambda^{C}{}_{B} + (\Lambda^{-1})^{A}{}_{C}\Lambda^{D}{}_{B}\omega^{C}{}_{D\mu}\,$. Therefore, if the affine connection, Eq.~(\ref{Weitzenboch connection}), holds in a specific frame $\theta^{A}{}_{\mu}$ then so does in another frame $\theta'^{A}{}_{\mu}=\Lambda^{A}{}_{B}\theta^{B}{}_{\mu}\,$. This also implies that we can identify the generic covariant derivative ``$\,\mathcal{D}\,$'' by that on spacetime ``$\,\nabla\,$'' as long as we use the affine connection, as mentioned in Ref.~\cite{Tomonari:2023ars}. In particular, we can always take so-called the Weitzenb\"{o}ck gauge~\cite{Adak:2005cd,Adak:2006rx,Adak:2008gd,Adak:2011ltj,Blagojevic:2023fys}
\begin{equation}
    \omega^{A}{}_{B\mu}=0\,.
\label{Weitzenboch gauge}
\end{equation}
Therefore, we have the so-called Weitzenb\"{o}ck connection as follows:
\begin{equation}
    \tilde{\Gamma}^{\rho}_{\mu\nu} = e_{A}{}^{\rho}\,\partial_{\mu}\theta^{A}{}_{\nu}\,,\quad \omega'^{A}{}_{B\mu} = (\Lambda^{-1})^{A}_{\ C}\partial_{\mu}\Lambda^{C}{}_{B}\,.
\label{Affine connection and connection 1-form in Weitzenboch gauge}
\end{equation}
In this specific gauge choice, the teleparallel condition is automatically satisfied, {\it i.e.} $\tilde{R}^{\sigma}{}_{\mu\nu\rho} = 2\partial_{[\nu}\tilde{\Gamma}^{\sigma}{}_{\rho]\mu} + 2\tilde{\Gamma}^{\sigma}{}_{[\nu|\lambda|}\tilde{\Gamma}^{\lambda}{}_{\rho]\mu} = 0\,$, which is independent in the choice of internal-space frames. While we have the relation
\begin{equation}
    \tilde{R}^{\sigma}{}_{\mu\nu\rho}=\overset{\circ}{R}{}^{\sigma}{}_{\mu\nu\rho}+2\overset{\circ}{\nabla}_{[\nu}N^{\sigma}{}_{\rho]\mu}+2N^{\sigma}{}_{[\nu|\lambda|}N^{\lambda}{}_{\rho]\mu}
\label{Generic curvature tensor}
\end{equation}
for a distorsion tensor $N^{\rho}{}_{\mu\nu}=\tilde{\Gamma}^{\rho}{}_{\mu\nu}-\overset{\circ}{\Gamma}{}^{\rho}{}_{\mu\nu}\,$, where $\overset{\circ}{\Gamma}{}^{\rho}{}_{\mu\nu}$ is the Levi-Civita connection, and ``$\,\overset{\circ}{\nabla}\,$'' denotes the covariant derivative with respect to the Levi-Civita connection. In MAG, the distorsion tensor is decomposed into the contorsion, $K^{\rho}{}_{\mu\nu}\,$, and the disformation, $L^{\rho}{}_{\mu\nu}\,$, as follows:
\begin{equation}
    N^{\rho}{}_{\mu\nu}=K^{\rho}{}_{\mu\nu}+L^{\rho}{}_{\mu\nu}\,,
\label{distorsion}
\end{equation}
where 
\begin{equation}
    K^{\rho}{}_{\mu\nu}=\frac{1}{2}T^{\rho}{}_{\mu\nu}+T_{(\mu\,\,\,\nu)}^{\,\,\,\,\,\rho}\,,\quad L^{\rho}{}_{\mu\nu}=\frac{1}{2}Q^{\rho}{}_{\mu\nu}-Q_{(\mu\,\,\,\nu)}^{\,\,\,\,\,\rho}\,,
\label{contorsion and disformation}
\end{equation}
respectively and, the torsion $T^{\rho}{}_{\mu\nu}$ and the non-metricity $Q^{\rho}{}_{\mu\nu}$ are defined by
\begin{equation}
    T^{\rho}{}_{\mu\nu}=e_{A}{}^{\rho}T^{A}{}_{\mu\nu} = 2e_{A}{}^{\rho}\partial_{[\mu}\theta^{A}{}_{\nu]} = -2\theta^{A}{}_{[\mu}\partial_{\nu]}e_{A}{}^{\rho}\,,\quad Q^{\rho}{}_{\mu\nu} = g^{\rho\lambda}\nabla_{\lambda}g_{\mu\nu}\,,
\label{torsion and nonmetricity}
\end{equation}
respectively, where we used the relation: Eq.~(\ref{relations btwn e and theta}) and the Weitzenb\"{o}ck gauge: Eq.~(\ref{Weitzenboch gauge}). In this gauge, a direct computation shows that the non-metricity automatically vanishes, provided that the internal-space metric is taken to be the Minkowskian metric according to the usual convention. Therefore, {\it in STEGR with the Weitzenb\"{o}ck gauge, we have to set the internal-space metric $\eta_{AB}$ as a more generic one such that the non-metricity arises from the internal-space structure.} One can find a similar case in massive gravity~\cite{deRham:2010kj,deRham:2011rn,Hassan:2011vm,Hinterbichler:2011tt,Arkani-Hamed:2002bjr,Dubovsky:2004sg,deRham:2014zqa}, as mentioned in Refs.~\cite{BeltranJimenez:2022azb,Hu:2023gui}. In addition, in STEGR, the torsion should vanish as a condition: $T^{\rho}{}_{\mu\nu} := 0\,$. In a special case, based on the local property $\left.\mathcal{M}\times\mathbb{R}^{n+1}\right|_{U}\simeq \left.T\mathcal{M}\right|_{U}\,$, we can decompose the co-frame field components as follows: $\theta^{A}{}_{\mu} = \partial_{\mu}\xi^{A}\,$, where $\xi^{A}$ are $n+1$ independent functions so-called the St\"{u}ckelberg fields~\cite{BeltranJimenez:2022azb,Hu:2023gui}. Then, the vanishing-torsion condition is automatically satisfied. It implies that we implicitly introduced some symmetry together with the independent functions $\xi^{A}$ that restricts the number (abbreviate it by denoting ``\#'' hereinafter) of $n(n+1)/2$ variables of the co-frame field components since the total number of those independent components under the local Lorentz symmetry is $(n+1)(n+2)/2\,$. Namely, this theory has the total number of $(n+1)(n+2)/2$ (\# of the independent components of the (co-)frame field) $\,-\,n(n+1)/2$ (\# of the restriction of some symmetry to the (co-)frame field) $\,=\,n+1$ independent variables for the (co-)frame field sector. This number is just the total number of the independent functions $\xi^{A}$. In the current paper, however, we do not utilize this decomposition and instead impose the vanishing-torsion property when either composing the Lagrangian density or composing the primary constraint densities, as will be shown in the sequel sections. Let us call the former and the latter theory ``{\it Formalism 1}'' and ``{\it Formalism 2}'', respectively. 

The co-frame field decomposition has a peculiar property. Namely, $\partial^{\mu}\xi_{A} = g^{\mu\nu}\,\partial_{\nu}\xi_{A}\,$ and Eq.~(\ref{relation between spacetime and internal space metric}) derive $\partial^{\mu}\xi_{A} = e_{I}{}^{\mu}\,e_{J}{}^{\nu}\,\eta^{IJ}\,\partial_{\nu}\xi_{A} = \partial^{\mu}\xi_{I}\,\partial^{\nu}\xi_{J}\,\eta^{IJ}\,\partial_{\nu}\xi_{A} = g^{\mu\rho}\,g^{\nu\lambda}\,\eta^{IJ}\,\partial_{\rho}\xi_{I}\,\partial_{\lambda}\xi_{J}\,\partial_{\nu}\xi_{A} = \cdots$ and so on. One would notice that this procedure not only never stops but also never removes the spacetime metric. Thus, we should treat it just as $\partial^{\mu}\xi_{A} = g^{\mu\nu}\,\partial_{\nu}\xi_{A}\,$. This would imply that the theory turns out to have a bi-metric structure that is composed of both the spacetime metric and the internal-space metric in an independent manner, which is, however, unfamiliar one in the well-known bi-metric theories of gravity~\cite{Rosen:1940zza,Rosen:1940zz,Hassan:2011zd,Schmidt-May:2015vnx}. In addition to this, in STEGR with the Weitzenb\"{o}ck gauge, remark that generically $\partial_{\mu}\xi^{A} = \eta^{AB}\,\partial_{\mu}\xi_{B} + \xi_{B}\,\partial_{\mu}\eta^{AB} \neq \eta^{AB}\,\partial_{\mu}\xi_{B}\,$. 
Let us call a theory that includes this unknown bi-metric characteristic {\it Formalism 3 } for the moment. In this work, we will not treat this theory. We will focus on unveiling internal symmetries in Formalism 1 and Formalism 2, leaving the investigation of Formalism 3 for a sequel paper.

\subsection{\label{02:02}Lagrangian density of STEGR in internal-space formulation}
Let us introduce the Lagrangian of STEGR, which is formulated in the internal-space. The vanishing-torsion property indicates so does for the contorsion, which is given in the first formula of Eq.~(\ref{contorsion and disformation}), and Eq.~(\ref{Generic curvature tensor}) together with Eq.~(\ref{distorsion}) and Eq.~(\ref{contorsion and disformation}) leads to
\begin{equation}
    \overset{\circ}{R} = \mathbb{Q} - \overset{\circ}{\nabla}_{\mu}\left(Q^{\mu} - \bar{Q}^{\mu}\right)\,,
\label{generic Ricci scalar in teleparallel}
\end{equation}
where we used the teleparallel condition and, $\mathbb{Q}$, $Q^{\mu}$, and $\bar{Q}^{\mu}$ are defined as follows:
\begin{equation}
\begin{split}
    &\mathbb{Q} = - \frac{1}{4}Q_{\rho\mu\nu}Q^{\rho\mu\nu} + \frac{1}{2}Q_{\rho\mu\nu}Q^{\mu\nu\rho} - \frac{1}{2}Q_{\mu}\bar{Q}^{\mu} + \frac{1}{4}Q_{\mu}Q^{\mu}\,,\\
    &Q^{\mu} = Q^{\mu\,\,\,\nu}_{\,\,\,\nu}\,,\\
    &\bar{Q}^{\mu} = Q^{\nu\,\,\,\mu}_{\,\,\,\nu}\,.
\end{split}
\label{bbQ Q^mu barQ^mu}
\end{equation}
Therefore, the generic STEGR Lagrangian density in Formalism 1 is given as follows:
\begin{equation}
    \mathcal{L}_{\rm generic\,Formalism\,1\,}(e_{A}{}^{\mu}\,,g_{\mu\nu}\,,\tau^{\mu\nu}{}_{\rho}) = \sqrt{-g}\,\mathbb{Q} + \tau^{\mu\nu}{}_{\rho}\,T^{\rho}{}_{\mu\nu} - \overset{\circ}{\nabla}_{\mu}\left[\sqrt{-g}\,\left(Q^{\mu} - \bar{Q}^{\mu}\right)\right]\,,
\label{generic STEGR Lagrangian in Formalism 1}
\end{equation}
where $g$ is the determinant of the spacetime metric components $g_{\mu\nu}$ and $\tau^{\mu\nu}{}_{\rho}$ is a set of Lagrange multipliers being anti-symmetric with respect to the upper two indices. In the second term of Eq.~(\ref{generic STEGR Lagrangian in Formalism 1}), the co-frame field components, $\theta^{A}{}_{\mu}\,$, are contained. However, notice that these components are not independent of the frame field components due to the relations given in Eq.~(\ref{relations btwn e and theta}). Therefore, the Lagrangian density depends only on $e_{A}{}^{\mu}$ but not on $\theta^{A}{}_{\mu}$. In addition, varying with respect to the multipliers $\tau^{\mu\nu}{}_{\rho}\,$, we obtain the vanishing-torsion property: $T^{\rho}{}_{\mu\nu} := 0$. This means that we regard these multipliers also as variables composing the configuration space of the theory. In the ordinary formulation of STEGR, however, the boundary term is dropped down on the ground of that the existence of a boundary term in a Lagrangian density does not change the equations of motion. Therefore, according to the standard formulation, we analyze the theory described by the following Lagrangian density:
\begin{equation}
\begin{split}
    &\mathcal{L}_{\rm Formalism\,1\,}(e_{A}{}^{\mu}\,,g_{\mu\nu}\,,\tau^{\mu\nu}{}_{\rho}) = \\
    &\quad\quad\quad\sqrt{-g}\,\left[ - \frac{1}{4}Q_{\rho\mu\nu}Q^{\rho\mu\nu} + \frac{1}{2}Q_{\rho\mu\nu}Q^{\mu\nu\rho} - \frac{1}{2}Q_{\mu}\bar{Q}^{\mu} + \frac{1}{4}Q_{\mu}Q^{\mu}\right]
    + \tau^{\mu\nu}{}_{\rho}\,T^{\rho}{}_{\mu\nu}\,.    
\end{split}
\label{STEGR Lagrangian in Formalism 1}
\end{equation}
Remark, here, that if we consider an extension of STEGR in a non-linear manner like $f(R)$-gravity~\cite{Buchdahl:1970ldb}, we cannot drop down the boundary term~\cite{Hu:2022anq,Tomonari:2023wcs,Bajardi:2024qbi}. Note that Eq.~(\ref{STEGR Lagrangian in Formalism 1}) is also considered in Ref.~\cite{Gomes:2022vrc} upon a physical motivation from the point of view of Noether charges with other possible formulations using Lagrange multipliers, in which four alternative formulations are investigated. (In detail, see Table I in Ref.~\cite{Gomes:2022vrc}.) In the current paper, we reformulate this theory in the internal-space below and investigate its gauge symmetry in Sec.~\ref{03}.

In the same manner, we can introduce the generic STEGR Lagrangian density in Formalism 2 just by dropping the second and third term in Eq.~(\ref{generic STEGR Lagrangian in Formalism 1}) and imposing the following primary constraint densities:
\begin{equation}
    \phi^{(1)\rho}{}_{\mu\nu} := e_{A}{}^{\rho}\,T^{A}{}_{\mu\nu} = 2\,e_{A}{}^{\rho}\partial_{[\mu}\theta^{A}{}_{\nu]} \approx 0\,.
\label{specific PC in Formalism 2 in spacetime}
\end{equation}
Remark that we do not utilize any decomposition of the frame field components mentioned in Sec.~\ref{02:01}. Under the satisfaction of this constraint density, the STEGR Lagrangian density in Formalism 2 is given as follows:
\begin{equation}
    \mathcal{L}_{\rm Formalism\,2\,}(e_{A}{}^{\mu}\,,g_{\mu\nu}) = \sqrt{-g}\,\left[ - \frac{1}{4}Q_{\rho\mu\nu}Q^{\rho\mu\nu} + \frac{1}{2}Q_{\rho\mu\nu}Q^{\mu\nu\rho} - \frac{1}{2}Q_{\mu}\bar{Q}^{\mu} + \frac{1}{4}Q_{\mu}Q^{\mu}\right]\,.
\label{STEGR Lagrangian in Formalism 2}
\end{equation}
The crucial point here is that, as mentioned in Sec.~\ref{02:01}, in STEGR theories the torsion does not automatically vanish. To overcome this issue, on one hand, the ordinary STEGR formulation employs the frame field decomposition explained in Sec.~\ref{02}. On the other hand, Formalism 1 and Formalism 2 impose the vanishing-torsion property on the Lagrangian density and in the primary constraint, respectively. Namely, the theories introduced above differ from the ordinary STEGR under the decomposition of the frame field components.

We can switch the formulation from that of metric on spacetime, {\it i.e.,} the metric formulation, to that of frame field and metric on internal-space, {\it i.e.,} the internal-space formulation. The non-metricity tensor given in Eq.~(\ref{torsion and nonmetricity}) is expressed as follows:\footnote{
Without the imposition of the Weitzenboeck gauge, the non-metricity in the internal-space is expressed as follows:
\begin{equation}
\begin{split}
        &Q^{\rho}{}_{\mu\nu} = e_{C}{}^{\rho}\,e_{D}{}^{\lambda}\,\theta^{A}{}_{\mu}\,\theta^{B}{}_{\nu}\,\eta^{CD}\,\partial_{\lambda}\eta_{AB} -2\,e_{D}{}^{\rho}\,\eta^{BD}\,\eta_{AC}\,\omega^{A}{}_{B[\mu}\,\theta^{C}{}_{\nu]}\,,\\
        &Q_{\rho\mu\nu} = \theta^{A}{}_{\mu}\,\theta^{B}{}_{\nu}\,\partial_{\rho}\eta_{AB} -2\,\theta^{B}{}_{\rho}\,\eta_{AC}\,\omega^{A}{}_{B[\mu}\,\theta^{C}{}_{\nu]}\,.
\end{split}
\label{non-metricity in terms of theta and eta without Weitzenboch gauge}
\end{equation}
Namely, the second term that is proportional to the connection 1-form components appears. 
}
\begin{equation}
    Q^{\rho}{}_{\mu\nu} = e_{C}{}^{\rho}\,e_{D}{}^{\lambda}\,\theta^{A}{}_{\mu}\,\theta^{B}{}_{\nu}\,\eta^{CD}\,\partial_{\lambda}\eta_{AB}\,\quad{\rm or}\,,\quad Q_{\rho\mu\nu} = \theta^{A}{}_{\mu}\theta^{B}{}_{\nu}\partial_{\rho}\eta_{AB}\,.
\label{non-metricity in terms of theta and eta}
\end{equation}
Then Eq.~(\ref{bbQ Q^mu barQ^mu}) can be expressed as follows:
\begin{equation}
\begin{split}
    &\mathbb{Q} = \frac{1}{2}\eta^{AB}\eta^{CD}\partial_{\mu}\eta_{A[B}\partial^{\mu}\eta_{C]D} + e_{A}{}^{\mu}e_{B}{}^{\nu}\eta^{AD}\eta^{BE}\eta^{CF}\partial_{\mu}\eta_{C[E}\partial_{\nu}\eta_{F]D}\,,\\
    &Q^{\mu} = e_{C}{}^{\mu}e_{D}{}^{\nu}\eta^{AB}\eta^{CD}\partial_{\nu}\eta_{AB}\,,\\
    &\bar{Q}^{\mu} = e_{C}{}^{\mu}e_{D}{}^{\nu}\eta^{AC}\eta^{BD}\partial_{\nu}\eta_{AB}\,.
\end{split}
\label{bbQ Q^mu barQ^mu in terms of theta and eta}
\end{equation}
Utilizing these expressions, the STEGR Lagrangian density in Formalism 1 becomes as follows:
\footnote{
If we take into account the boundary term then the Lagrangian density becomes as follows:
\begin{equation}
\begin{split}
    &\mathcal{L}_{\rm generic\,internal\,Formalism\,1}(e_{A}{}^{\mu}\,,\eta_{AB}\,,\tau^{AB}{}_{\mu}) =\\
    &\quad\quad\quad\frac{1}{2}\,e\,\sqrt{-\eta}\,e_{A}{}^{\mu}\,e_{B}{}^{\nu}\,\eta^{ABCDEF}\,\partial_{\mu}\eta_{C[D}\partial_{\nu}\eta_{E]F} 
    - 2\,\tau^{AB}{}_{\mu}\,e_{[A}{}^{\nu}\,\partial_{\nu}e_{B]}{}^{\mu}\\
    &\quad\quad\quad- \overset{\circ}{\nabla}_{\mu}\left[2\,e\,\sqrt{-\eta}\,e_{C}{}^{\mu}\,e_{D}{}^{\nu}\,\eta^{A[B}\eta^{C]D}\partial_{\nu}\eta_{AB}\,\right]\,.    
\end{split}
\label{STEGR Lagrangian in terms of e and eta with boundary term}
\end{equation}
Namely, the third term appears. In Formalism 2, of course, the second term is absent.
}
\begin{equation}
\begin{split}
    &\mathcal{L}_{\rm\,internal\,Formalism\,1\,}(e_{A}{}^{\mu}\,,\eta_{AB}\,,\tau^{AB}{}_{\mu}) =\\
    &\quad\quad\quad\frac{1}{2}\,e\,\sqrt{-\eta}\,e_{A}{}^{\mu}\,e_{B}{}^{\nu}\,\eta^{ABCDEF}\,\partial_{\mu}\eta_{C[D}\partial_{\nu}\eta_{E]F} 
        - 2\,\tau^{AB}{}_{\mu}\,e_{[A|}{}^{\nu}\,\partial_{\nu}e_{|B]}{}^{\mu}
\end{split}
\label{STEGR Lagrangian in terms of e and eta in Formalism 1}
\end{equation}
where we set the auxiliary variable $\tau^{AB}{}_{\mu}$ and the super-metric $\eta^{ABCDEF}$ by
\begin{equation}
    \tau^{AB}{}_{\rho} := \tau^{\mu\nu}{}_{\rho}\,\theta^{A}{}_{\mu}\,\theta^{B}{}_{\nu} = -\tau^{BA}{}_{\rho}\,,\quad\eta^{ABCDEF} := \eta^{AB}\eta^{CD}\eta^{EF} + 2\eta^{AF}\eta^{BD}\eta^{CE}\,,
\label{tau and eta}
\end{equation}
respectively, and $e := \epsilon^{A_{0}A_{1}A_{2}A_{3}\,\cdots\,A_{n}} e_{A_{0}}{}^{0}e_{A_{1}}{}^{1}e_{A_{2}}{}^{2}e_{A_{3}}{}^{3}\,\cdots\,e_{A_{n}}{}^{n}$ and $\eta := \epsilon^{\mu_{0}\mu_{1}\mu_{2}\mu_{3}\,\cdots\,\mu_{n}}\eta_{0\mu_{0}}\eta_{1\mu_{1}}\eta_{2\mu_{2}}\eta_{3\mu_{3}}\,\cdots\,\eta_{n\mu_{n}}$ are the determinant of the frame field components and the internal-space metric, respectively, where $\epsilon^{\cdots}$ is the Levi-Civita anti-symmetric symbol. Varying with respect to $\tau^{AB}{}_{\mu}$ and using Eq.~(\ref{relations btwn e and theta}), we obtain the vanishing-torsion property. Therefore, we can replace the auxiliary variables $\tau^{\mu\nu}{}_{\rho}$ by $\tau^{AB}{}_{\mu}\,$. In Formalism 2, the Lagrangian density is given by
\begin{equation}
        \mathcal{L}_{\rm\,internal\,Formalism\,2\,}(e_{A}{}^{\mu}\,,\eta_{AB}) = \frac{1}{2}\,e\,\sqrt{-\eta}\,e_{A}{}^{\mu}\,e_{B}{}^{\nu}\,\eta^{ABCDEF}\,\partial_{\mu}\eta_{C[D|}\partial_{\nu}\eta_{|E]F}\,. 
\label{STEGR Lagrangian in terms of e and eta in Formalism 2}
\end{equation}
The primary constraint density, Eq.~(\ref{specific PC in Formalism 2 in spacetime}), is expressed by
\begin{equation}
    \tilde{\phi}^{(1)A}{}_{BC} := \theta^{A}{}_{\rho}\,e_{B}{}^{\mu}\,e_{C}{}^{\nu}\,\phi^{(1)\rho}{}_{\mu\nu} = -\,2\,\theta^{A}{}_{\nu}\,e_{[B|}{}^{\mu}\,\partial_{\mu}e_{|C]}{}^{\nu} \approx 0\,,
\label{specific PC in Formalism 2 in internal-space}
\end{equation}
or equivalently, on a local region in which the frame field is invertible, the above equation turns into
\begin{equation}
    \phi^{(1)\mu}{}_{BC} := \,e_{B}{}^{\nu}\,e_{C}{}^{\rho}\,\phi^{(1)\mu}{}_{\nu\rho} = -\,2\,e_{[B|}{}^{\rho}\,\partial_{\rho}e_{|C]}{}^{\mu} \approx 0\,,
\label{specific PC in Formalism 2 only in e}
\end{equation}
where we used Eq.~(\ref{relations btwn e and theta}). The total number of the components of the constraint density is $n(n + 1)^{2}/2$, and this number is less than the total number of the configuration variables: $(n + 1)(3n + 4)/2$. We investigate these Lagrangian in the current article. Let us call the theories provided by Eq.~(\ref{STEGR Lagrangian in terms of e and eta in Formalism 1}) and Eq.~(\ref{STEGR Lagrangian in terms of e and eta in Formalism 2}) ``{\it internal STEGR}'' in Formalism 1 and Formalism 2, respectively, to distinguish from the original STEGR theories.

For the internal STEGR in Formalism 2, we can consider a special case by imposing the frame field decomposition for which let us call the internal STEGR in Formalism 3. The Lagrangian density is given as follows:
\begin{equation}
\begin{split}
        &\mathcal{L}_{\rm\,internal\,Formalism\,3\,}(\xi_{A}{}^{}\,,g^{\mu\nu}\,,\eta_{AB}) =\\
        &\quad\quad\quad\quad\frac{1}{2}\,\xi\,\sqrt{-\eta}\,\eta^{ABCDEF}\,g^{\mu\alpha}\,g^{\nu\beta}\,\partial_{\alpha}\xi_{A}{}^{}\,\partial_{\beta}\xi_{B}{}^{}\,\partial_{\mu}\eta_{C[D|}\partial_{\nu}\eta_{|E]F}\,,
\end{split}
\label{STEGR Lagrangian in terms of xi and eta in Formalism 3}
\end{equation}
where $\xi$ is the determinant of $\partial^{\mu}\xi_{A} = g^{\mu\nu}\,\partial_{\nu}\xi_{A}\,$. As mentioned in Sec.~\ref{02:01}, in the frame filed decomposition $e_{A}{}^{\mu} = \partial^{\mu}{}_{}\xi_{A}\,$, the vanishing-torsion property is automatically satisfied without any additional conditions. The St\"{u}ckelberg fields $\xi_{A} = \eta_{AB}\,\xi^{B}$ has its own {\it global} symmetry for a pure group action. Namely, for a constant element $\tilde{\Lambda}^{A}{}_{B}\,\in\,\tilde{G}\,$, the St\"{u}ckelberg fields $\xi^{A}$ is transformed by $\xi'^{A} = \tilde{\Lambda}^{A}{}_{B}\,\xi^{B}\,$, and the theory is invariant under the satisfaction of the following transformation to the internal-space metric:\footnote{
We can extend the transformation into the affine transformation as follows:
\begin{equation}
    \xi'^{A} = \tilde{\Lambda}^{A}{}_{B}\,\xi^{B} + \zeta^{A}\,,
\label{}
\end{equation}
where $\zeta^{A}$ is a $(n+1)$-vector in the internal-space. A simple way to make the theory invariant under this transformation is $\xi^{A}\rightarrow \xi^{A} + \eta^{AB}\,\zeta^{B}$ and $\eta^{AB} \rightarrow \eta^{AB}$.
}
\begin{equation}
    \eta_{AB} = \eta'_{IJ}\,\tilde{\Lambda}^{I}{}_{A}\,\tilde{\Lambda}^{J}{}_{B}\,.
\label{Transformation law of eta in Struckelberg fields transformation}
\end{equation}
Thus, the internal STEGR in Formalism 3 has not only {\it local} internal-space symmetries but also {\it global} internal-space symmetries. Notice that this symmetry reduces the total number of the independent components of the frame field components from $(n + 1)(n + 2)/2$ to $n + 1$ and this number matches to that of the St\"{u}ckelberg fields and of the coordinates at each point in spacetime. This motivates further simplification as follows. Since a local region where the structure $\left.\mathcal{M}\times\mathbb{R}^{n+1}\right|_{U}\simeq \left.T\mathcal{M}\right|_{U}$ holds, we can identify $\xi^{A}$ as $\xi^{\mu}\,$, and then $\xi^{\mu} := x^{\mu}\,$~\cite{Tomonari:2023ars,Tomonari:2023wcs}. This property also allows us to identify the internal-space indices and the spacetime indices. In this case, the frame field components satisfy $e_{A}{}^{\mu} := e_{\nu}{}^{\mu} = \delta_{\nu}{}^{\mu}\,$, or equivalently, $\theta^{A}{}_{\mu} := \theta^{\nu}{}_{\mu} = \delta^{\nu}{}_{\mu}\,$, and then $\eta_{AB} = e_{A}{}^{\mu}\,e_{B}{}^{\nu}\,g_{\mu\nu} = g_{AB}\,$ and $g_{\mu\nu} = \theta^{A}{}_{\mu}\,\theta^{B}{}_{\nu}\,\eta_{AB} = \eta_{\mu\nu}\,$. This suggests that the local region $U$ in $\left.\mathcal{M}\times\mathbb{R}^{n+1}\right|_{U}\simeq \left.T\mathcal{M}\right|_{U}$ should be extended to the entire spacetime manifold $\mathcal{M}$: $U = \mathcal{M}\,$. Namely, the local property turns into the global property. This is nothing but the so-called coincident gauge~\cite{BeltranJimenez:2022azb,Hu:2023gui}. In this gauge, STEGR in the internal formulation coincides with that in the ordinary spacetime formulation. In this work, we do not adopt the coincident gauge to establish the internal formulation of STEGR. Also, as mentioned in Sec.~\ref{02:01}, Formalism 3, on one hand, shows an unfamiliar bi-metric structure. On the other hand, theories of STEGR in the coincident gauge completely lose the structure. We will not discuss the internal STEGR in Formalism 3 in the current paper and leave it for a sequel paper since Formalism 3 has its own interesting constraint structure differing from that of Formalism 1 and Formalism 2, focusing on the investigation of that in Formalism 1 and Formalism 2.

\section{\label{03}Internal STEGR in Formalism 1}
\subsection{\label{03:01}Canonical momenta and Primary constraints}
The Lagrangian density of the internal STEGR in Formalism 1 was introduced as Eq.~(\ref{STEGR Lagrangian in terms of e and eta in Formalism 1}) in the previous subsection. The configuration space $\mathcal{Q}_{1}$ is coordinated by the three set of variables: $e_{A}{}^{\mu}\,$, $\eta_{AB}\,$, and $\tau^{AB}{}_{\mu}\,$. Let us denote it as $\mathcal{Q}_{1} = \left<e_{A}{}^{\mu}\,,\eta_{AB}\,,\tau^{AB}{}_{\mu}\right>\,$. Thus, the velocity phase-space of the theory is given by the tangent bundle of $\mathcal{Q}_{1}\,$, {\it i.e.,} $T\mathcal{Q}_{1} = \left<e_{A}{}^{\mu}\,,\eta_{AB}\,,\tau^{AB}{}_{\mu}\,;\,\dot{e}_{A}{}^{\mu}\,,\dot{\eta}_{AB}\,,\dot{\tau}^{AB}{}_{\mu}\right>\,$, and the phase-space is nothing but its dual-bundle, {\it i.e.,} $T^{*}\mathcal{Q}_{1} = \left<e_{A}{}^{\mu}\,,\eta_{AB}\,,\tau^{AB}{}_{\mu}\,;\,\pi^{A}{}_{\mu}\,,\pi^{AB}\,,\pi_{AB}{}^{\mu}\right>\,$, where $\pi^{A}{}_{\mu}\,$, $\pi^{AB}\,$, and $\pi_{AB}{}^{\mu}$ are canonical momentum variables with respect to each configuration variable, {\it i.e.,} $e_{A}{}^{\mu}\,$, $\eta_{AB}$ and $\tau^{AB}{}_{\mu}$ given as follows: 
\begin{equation}
    \pi^{A}{}_{\mu} = \frac{\delta\mathcal{L}_{\rm internal\,Formalism\,1}}{\delta\dot{e}_{A}{}^{\mu}} = -2\tau^{IA}{}_{\mu}\,e_{I}{}^{0}\,,
\label{CM wrt e in Formalism 1}
\end{equation}
\begin{equation}
    \pi^{AB} = \frac{\delta \mathcal{L}_{\rm internal\,Formalism\,1}}{\delta \dot{\eta}_{AB}} = D^{ABEF}\,\dot{\eta}_{EF} + \frac{1}{2}\,e\,\sqrt{-\eta}\,e_{C}{}^{0}\,e_{D}{}^{i}\,\partial_{i}\eta_{EF}\,\tilde{\eta}^{CDABEF}\,,
\label{CM wrt eta in Formalism 1}
\end{equation}
where $\tilde{\eta}^{ABCDEF}$ and $D^{ABCD}$ are defined by
\begin{equation}
\begin{split}
    &\tilde{\eta}^{ABCDEF} := 2\eta^{A[B}\,\eta^{E]F}\,\eta^{CD} + \eta^{AE}\,\eta^{B(C}\,\eta^{D)F} - \eta^{AB}\,\eta^{E(C}\,\eta^{D)F}\\
    &\quad\quad\quad\quad\quad\quad\quad + \eta^{B(C}\,\eta^{D)E}\,\eta^{AF} - \eta^{A(C}\,\eta^{D)B}\,\eta^{EF}\,,\\
    &D^{ABCD} := \frac{\delta \pi^{AB}}{\delta \dot{\eta}_{CD}} = \frac{1}{2}\,e\,\sqrt{-\eta}\,e_{I}{}^{0}\,e_{J}{}^{0}\,\tilde{\eta}^{IJABCD}\,,
\end{split}
\label{tilde_eta and D}
\end{equation}
and
\begin{equation}
    \pi_{AB}{}^{\mu} = \frac{\delta \mathcal{L}_{\rm internal\,Formalism\,1}}{\delta \dot{\tau}^{AB}{}_{\mu}} = 0\,,
\label{CM wrt tau}
\end{equation}
respectively. Fundamental PB-algebra is introduced as follows:
\begin{equation}
\begin{split}
    &\{e_{A}{}^{\mu}(t\,,\vec{x})\,,\pi^{B}{}_{\nu}(t\,,\vec{y})\} = \delta_{A}{}^{B}\,\delta^{\mu}{}_{\nu}\,\delta^{(n)}(\vec{x} - \vec{y})\,,\\
    &\{\eta_{AB}(t\,,\vec{x})\,,\pi^{CD}(t\,,\vec{y})\} = \delta^{C}{}_{(A}\,\delta^{D}{}_{B)}\,\delta^{(n)}(\vec{x} - \vec{y})\,,\\
    &\{\tau^{AB}{}_{\mu}(t\,,\vec{x})\,,\pi_{CD}{}^{\nu}(t\,,\vec{y})\} = \delta^{A}{}_{[C}\,\delta^{B}{}_{D]}\,\delta_{\mu}{}^{\nu}\,\delta^{(n)}(\vec{x} - \vec{y})\,.
\end{split}
\label{Fundamental PB-algebra in Formalism 1}
\end{equation}
The Lagrangian density, Eq.~(\ref{STEGR Lagrangian in terms of e and eta in Formalism 1}), is defined on the velocity phase-space $T\mathcal{Q}_{1}$. Performing the Legendre transformation from $T\mathcal{Q}_{1}$ to the phase-space $T^{*}\mathcal{Q}_{1}$, we can switch the formulation of the theory from Lagrangian formulation to Hamiltonian one by introducing the so-called total-Hamiltonian. To do this, we have to unveil all primary constraint densities of the theory since the transformation should be taken place under the satisfaction of all the primary constraint densities. In addition, it would be expected that internal symmetries appear in the PB-algebra among the primary constraint densities since the transformation in the internal-space should be performed while holding the internal symmetries at each spacetime point as a necessary condition. In the current paper, we focus only on revealing possible internal symmetries.\footnote{
For spacetime symmetries, we should consider the Legendre transformation not at a point but on an open set of spacetime. It implies that a specific PB-algebra of spacetime symmetries appears among secondary or higher-order constraint densities. 
} Other symmetries including spacetime symmetries such as diffeomorphism symmetry would be investigated in the sequel papers on the Dirac-Bergmann analysis of our theory since the symmetries should be represented in the PB-algebras among secondary constraint densities~\cite{Dirac:1950pj,Dirac:1958sq,Bergmann:1949zz,BergmannBrunings1949,Bergmann1950,Anderson:1951ta}. 

On one hand, Eq.~(\ref{CM wrt e in Formalism 1}) and Eq.~(\ref{CM wrt tau}) do not contain any velocity variables and, therefore, lead directly to the following primary constraint densities:
\begin{equation}
    \phi^{(1)A}_{\,\,\,\,\,\,\,\,\,\,\,\,\mu} := \pi^{A}{}_{\mu} + 2\tau^{IA}{}_{\mu}\,e_{I}{}^{0} \approx 0\,,\quad \phi^{(1)\,\,\,\,\,\,\,\,\,\,\mu}_{\,\,\,\,\,\,\,AB} := \pi_{AB}{}^{\mu} \approx 0\,,
\label{PC wrt tau and e}
\end{equation}
respectively. To investigate internal symmetries in Sec.~\ref{03:02}, the first formula in Eq.~(\ref{PC wrt tau and e}) should be pulled back to the internal-space by the frame field as follows:
\begin{equation}
    \tilde{\phi}^{(1)AB} := -\,\phi^{(1)A}{}_{\rho}\,\eta^{BI}\,e_{I}{}^{\rho} \approx 0\,
\label{modified PC wrt e in Formalism 1}
\end{equation}
and decomposed it into the anti-symmetric and symmetric part as follows:
\begin{equation}
    \tilde{\phi}^{(1aS)AB} := 2\,\tilde{\phi}^{(1)[AB]} = -\,2\,\phi^{(1)[A}{}_{\rho}\,\eta^{B]I}\,e_{I}{}^{\rho} \approx 0\,
\label{modified PC1 wrt e in Formalism 1}
\end{equation}
and
\begin{equation}
    \tilde{\phi}^{(1S)AB} := 2\,\tilde{\phi}^{(1)(AB)} := -\,2\,\phi^{(1)(A}{}_{\rho}\,\eta^{B)I}\,e_{I}{}^{\rho} \approx 0\,,
\label{modified PC2 wrt e in Formalism 1}
\end{equation}
respectively. The total number of the components of these primary constraint densities, {\it i.e.,} the second formula $\phi^{(1)\,\,\,\,\,\,\,\,\,\,\mu}_{\,\,\,\,\,\,\,AB} \approx 0$ given in Eq.~(\ref{PC wrt tau and e}), $\tilde{\phi}^{(1aS)AB} \approx 0$ given in Eq.~(\ref{modified PC1 wrt e in Formalism 1}), and $\tilde{\phi}^{(1S)AB} \approx 0$ given in Eq.~(\ref{modified PC2 wrt e in Formalism 1}) are $n(n + 1)/2 + (n + 1)(n + 2)/2 + n(n + 1)^{2}/2 = (n + 1)^{2}(n + 2)/2$ and, of course, this number coincides with the upper bound of that of primary constraint densities with respect to $e_{A}{}^{\mu}$ and $\tau^{AB}{}_{\mu}\,$. 

On the other hand, Eq.~(\ref{CM wrt eta in Formalism 1}) provides primary constraint densities with respect to $\eta^{AB}$ up to the number of $(n + 1)(n + 2)/2\,$. To find them, we rewrite Eq.~(\ref{CM wrt eta in Formalism 1}) as follows:
\begin{equation}
    D^{ABIJ}\,\dot{\eta}_{IJ} = \pi^{AB} - \frac{1}{2}\,e\,\sqrt{-\eta}\,e_{K}{}^{0}\,e_{L}{}^{i}\,\partial_{i}\eta_{IJ}\,\tilde{\eta}^{KLABIJ}\,.
\label{deformed CM wrt eta}
\end{equation}
This equation implies that quantity $C_{AB}{}^{\cdots}$ with satisfying the property $C_{AB}{}^{\cdots}\,D^{ABIJ}\,\dot{\eta}_{IJ} = 0$ provides primary constraint densities  
\begin{equation}
    \phi^{\cdots} := C_{AB}{}^{\cdots}\,\left[\pi^{AB} - \frac{1}{2}\,e\,\sqrt{-\eta}\,e_{I}{}^{0}\,e_{J}{}^{i}\,\partial_{i}\eta_{KL}\,\tilde{\eta}^{IJABKL}\right] \approx 0\,,
\label{model of PC wrt eta}
\end{equation}
where ``$\cdots$'' denotes dummy indices. For such $C_{AB}{}^{\cdots}\,$, the most simple choice may be an anti-symmetric tensor and we find the following quantity:
\begin{equation}
    C_{AB\mu} := \eta_{AC}\,\eta_{BD}\,\tau^{CD}{}_{\mu} = - C_{BA\mu}\,.
\label{ingredients of PC wrt eta}
\end{equation}
This composition makes sense since primary constraint densities with respect to the internal-space metric should consist of the metric itself, and other possible variables to compose the constraint densities should be quantities that have both the indices of the internal-space and the spacetime, in which the internal-space indices should further be completely anti-symmetric. In addition to this primary constraint density, the second formula in Eq.~(\ref{PC wrt tau and e}) provides a primary constraint density which contains the spacetime index only. Therefore, primary constraint densities in this sector are given as follows:
\begin{equation}
    \phi^{(1)}{}_{\mu} := C_{AB\mu}\,\pi^{AB} \approx 0\,,\quad 
    \phi^{(1)\mu} := \phi^{(1)\,\,\,\,\,\,\,\,\,\,\mu}_{\,\,\,\,\,\,\,AB}\,\pi^{AB} \approx 0\,.
\label{PC wrt eta}
\end{equation}
The total number of the components of these primary constraint densities is $2(n + 1)$, and this number is less than that of the upper bound: $(n + 1)(n + 2)/2$. To investigate the frame symmetry in Sec.~\ref{03:02}, the first formula should be pulled back to the internal-space by the frame field as follows:
\begin{equation}
    \tilde{\phi}^{(1)}{}_{A} := e_{A}{}^{\rho}\,\phi^{(1)}{}_{\rho} = e_{A}{}^{\rho}\,C_{IJ\rho}\,\pi^{IJ} \approx 0\,,
\label{modified PC wrt eta 2}
\end{equation}
Note that the second formula cannot perform the pull-back manipulation since the spacetime index is equipped as upper one. To pull back such a quantity, we have to use the inverse frame field components, but in our theory the configuration space does not contain the inverse components. All quantities in our theory have to contain only the frame field components $e_{A}{}^{\mu}\,$, the internal-space metric $\eta_{AB}$, the auxiliary field $\tau^{AB}{}_{\mu}\,$, and those canonical momenta $\pi^{A}{}_{\mu}\,$, $\pi^{AB}\,$, and $\pi_{AB}{}^{\mu}\,$. Notice, finally, that it is expected that the primary constraint densities only with the internal-space indices, {\it i.e.,} Eq.~(\ref{modified PC wrt e in Formalism 1}), or equivalently, Eq.~(\ref{modified PC1 wrt e in Formalism 1}) and Eq.~(\ref{modified PC2 wrt e in Formalism 1}), and Eq.~(\ref{modified PC wrt eta 2}), provide internal symmetries. Once we find all first-class constraints of a theory, we can compose a generator of gauge transformation based on these constraints, and we can investigate symmetries of the theory~\cite{Sugano:1986xb,Sugano:1989rq,Sugano:1991ke,Sugano:1991kd,Sugano:1991ir}. In the next subsection~\ref{03:02}, we reveal possible symmetries in the internal STEGR in Formalism 1.

\subsection{\label{03:02}PB-algebras and Possible Symmetries}
We calculate the PB-algebra of the primary constraint densities, $\tilde{\phi}^{(1)AB} \approx 0\,$, $\tilde{\phi}^{(1)}{}_{A} \approx 0\,$, and $\phi^{(1)\mu} \approx 0\,$, in the following three steps. First, the PB-algebra restricts the entire phase-space that contains the canonical momenta $\pi^{A}{}_{\mu}\,$. Second, the PB-algebra restricts the entire phase-space that contains the canonical momenta $\pi^{AB}$. Third, the PB-algebra of the primary constraints that contains all the canonical momenta. Let us call them $\pi^{A}{}_{\mu}$-sector, $(\pi^{AB}\,,\pi_{AB}{}^{\mu})$-sector, and $(\pi^{A}{}_{\mu}\,,\pi_{AB}{}^{\mu}\,,\pi^{AB})$-sector (the entire sector), respectively.

The PB-algebra of $\pi^{A}{}_{\mu}$-sector is calculated as follows (See Appendix~\ref{App:01} for the derivation):
\begin{equation}
    \{\tilde{\phi}^{(1)AB}(t\,,\vec{x})\,,\tilde{\phi}^{(1)CD}(t\,,\vec{x})\} = 
    -\,4\,\eta^{BI}\,\eta^{DJ}\,e_{(I}{}^{\rho}\,e_{J)}{}^{0}\,\tau^{AC}{}_{\rho} + \eta^{AD}\,\tilde{\phi}^{(1)CB} - \eta^{CB}\,\tilde{\phi}^{(1)AD}\,.
\label{PB btw tilde phi_AB in Formalism 1}
\end{equation}
In particular, using the decompositions given in Eq.(\ref{modified PC1 wrt e in Formalism 1}) and Eq.(\ref{modified PC2 wrt e in Formalism 1}), the PB-algebra is split into the following three PB-algebras:
\begin{equation}
\begin{split}
    &\{\tilde{\phi}^{(1aS)AB}(t\,,\vec{x})\,,\tilde{\phi}^{(1aS)CD}(t\,,\vec{y})\} \\
    & = \quad \left[\eta^{BD}\,\tilde{\phi}^{(1aS)AC} + \eta^{AC}\,\tilde{\phi}^{(1aS)BD} - \eta^{BC}\,\tilde{\phi}^{(1aS)AD} - \eta^{AD}\,\tilde{\phi}^{(1aS)BC}\right]\,\delta^{(n)}(\vec{x} - \vec{y})\\
    &\quad\quad + 4\,e_{(I}{}^{\rho}\,e_{J)}{}^{0}\,\left[-\,\eta^{BI}\,\eta^{DJ}\,\tau^{AC}{}_{\rho} -\,\eta^{AI}\,\eta^{CJ}\,\tau^{BD}{}_{\rho}\right.\\ 
    &\quad\quad\quad\quad\quad\quad\quad\quad\quad\quad\quad\quad\quad\quad\quad\quad \left.+ \eta^{BI}\,\eta^{CJ}\,\tau^{AD}{}_{\rho} + \eta^{AI}\,\eta^{DJ}\,\tau^{BC}{}_{\rho}\right]\,\delta^{(n)}(\vec{x} - \vec{y})\,,\\
\end{split}
\label{PB btw tilde phi_AB in anti-symmetric in Formalism 1}
\end{equation}
\begin{equation}
\begin{split}
    &\{\tilde{\phi}^{(1aS)AB}(t\,,\vec{x})\,,\tilde{\phi}^{(1S)CD}(t\,,\vec{y})\} \,\\
    & = \quad \left[\eta^{BD}\,\tilde{\phi}^{(1S)AC} - \eta^{AC}\,\tilde{\phi}^{(1S)BD} +\,\eta^{BC}\,\tilde{\phi}^{(1S)AD} - \eta^{AD}\,\tilde{\phi}^{(1S)BC}\right]\,\delta^{(n)}(\vec{x} - \vec{y})\, \\
    &\quad\quad -\,4\,e_{(I}{}^{\rho}\,e_{J)}{}^{0}\,\left[\eta^{BI}\,\eta^{DJ}\,\tau^{AC}{}_{\rho} -\,\eta^{AI}\,\eta^{CJ}\,\tau^{BD}{}_{\rho}\right.\\
    &\quad\quad\quad\quad\quad\quad\quad\quad\quad\quad\quad\quad\quad\quad\quad\quad \left.+ \eta^{BI}\,\eta^{CJ}\,\tau^{AD}{}_{\rho} -\,\eta^{AI}\,\eta^{DJ}\,\tau^{BC}{}_{\rho}\right]\,\delta^{(n)}(\vec{x} - \vec{y})\,,
\end{split}
\label{PB btw tilde phi_AB in anti-symmetric and symmetric in Formalism 1}
\end{equation}
and
\begin{equation}
\begin{split}
    &\{\tilde{\phi}^{(1S)AB}(t\,,\vec{x})\,,\tilde{\phi}^{(1S)CD}(t\,,\vec{y})\} \,\\
    & = \quad \left[\eta^{BD}\,\tilde{\phi}^{(1aS)AC} + \eta^{AC}\,\tilde{\phi}^{(1aS)BD} + \eta^{BC}\,\tilde{\phi}^{(1aS)AD} + \eta^{AD}\,\tilde{\phi}^{(1aS)BC}\right]\,\delta^{(n)}(\vec{x} - \vec{y})\, \,,\\
    &\quad\quad -\,4\,e_{(I}{}^{\rho}\,e_{J)}{}^{0}\,\left[\eta^{BI}\,\eta^{DJ}\,\tau^{AC}{}_{\rho} + \eta^{AI}\,\eta^{CJ}\,\tau^{BD}{}_{\rho}\right. \\
    &\quad\quad\quad\quad\quad\quad\quad\quad\quad\quad\quad\quad\quad\quad\quad\quad \left.+ \eta^{BI}\,\eta^{CJ}\,\tau^{AD}{}_{\rho} + \eta^{AI}\,\eta^{DJ}\,\tau^{BC}{}_{\rho}\right]\,\delta^{(n)}(\vec{x} - \vec{y})\,.\\
\end{split}
\label{PB btw tilde phi_AB in symmetric in Formalism 1}
\end{equation}
Therefore, Eq.~(\ref{modified PC wrt e in Formalism 1}), or equivalently, Eq.~(\ref{modified PC1 wrt e in Formalism 1}) and Eq.~(\ref{modified PC2 wrt e in Formalism 1}), is classified as second-class due to the existence of the terms containing $\tau^{AB}{}_{\mu}\,$. Notice that these PB-algebras indicate that the existence of the auxiliary variable $\tau^{AB}{}_{\mu}$ breaks the internal symmetry. Namely, the second term in the Lagrangian density of Eq.~(\ref{STEGR Lagrangian in terms of e and eta in Formalism 1}) causes the violation of the internal symmetry. In fact, Eq.~(\ref{PB btw tilde phi_AB in Formalism 1}), or equivalently, Eq.~(\ref{PB btw tilde phi_AB in anti-symmetric in Formalism 1}), Eq.~(\ref{PB btw tilde phi_AB in anti-symmetric and symmetric in Formalism 1}), and Eq.~(\ref{PB btw tilde phi_AB in symmetric in Formalism 1}) contains the sub-algebra of the internal symmetry such as the local Lorentz symmetry. Namely, if the configuration variable $\tau^{AB}{}_{\mu}$ were removed, then these PB-algebras form a closed algebra. In particular, Eq.~(\ref{PB btw tilde phi_AB in anti-symmetric in Formalism 1}) in this case is noting but the Lorentz algebra. In this point, we will mention again in the last paragraph of this subsection.

The PB-algebra of $(\pi^{AB}\,,\pi_{AB}{}^{\mu})$-sector is calculated as follows (See Appendix~\ref{App:01} for the derivation):
\begin{equation}
\begin{split}
    &\{\tilde{\phi}^{(1)}{}_{A}(t\,,\vec{x})\,,\tilde{\phi}^{(1)}{}_{B}(t\,,\vec{y})\} = 0\,,\\
    &\{\tilde{\phi}^{(1)}{}_{A}(t\,,\vec{x})\,,\phi^{(1)\mu}(t\,,\vec{y})\} = 0\,,\\
    &\{\phi^{(1)\mu}(t\,,\vec{x})\,,\phi^{(1)\nu}(t\,,\vec{y})\} = 0\,.
\end{split}
\label{PB btw tilde phi_AB and hat phi_AB}
\end{equation}
Namely, all the constraint densities are commutative as the strong equality. 

The PB-algebra of $(\pi^{A}{}_{\mu}\,,\pi_{AB}{}^{\mu}\,,\pi^{AB})$-sector (the entire sector) is calculated as follows (See Appendix~\ref{App:01} for the derivation):
\begin{equation}
\begin{split}
    &\{\tilde{\phi}^{AB}(t\,,\vec{x})\,,\tilde{\phi}^{(1)}{}_{C}(t\,,\vec{y})\} = \delta^{A}{}_{C}\,\eta^{BI}\,\tilde{\phi}^{(1)}{}_{I}\,\delta^{(n)}(\vec{x} - \vec{y})\,,\\
    &\{\tilde{\phi}^{AB}(t\,,\vec{x})\,,\phi^{(1)\mu}(t\,,\vec{y})\} = 0\,.
\end{split}
\label{PB btw tilde phi^AB, tilde phi_A, and phi_AB-mu}
\end{equation}
Equivalently, splitting $\tilde{\phi}^{(1)AB}{}_{} \approx 0$ into the anti-symmetric and symmetric part, the first PB-algebra above equations becomes as follows:
\begin{equation}
\begin{split}
    &\{\tilde{\phi}^{(1aS)AB}(t\,,\vec{x})\,,\tilde{\phi}^{(1)}{}_{C}(t\,,\vec{y})\} = 2\,\delta^{[A}{}_{C}\,\eta^{B]I}\,\tilde{\phi}^{(1)}{}_{I}\,\delta^{(n)}(\vec{x} - \vec{y})\,,\\
    &\{\tilde{\phi}^{(1aS)AB}(t\,,\vec{x})\,,\phi^{(1)\mu}(t\,,\vec{y})\} = 0\,,\\
    &\{\tilde{\phi}^{(1S)AB}(t\,,\vec{x})\,,\tilde{\phi}^{(1)}{}_{C}(t\,,\vec{y})\} = 2\,\delta^{(A}{}_{C}\,\eta^{B)I}\,\tilde{\phi}^{(1)}{}_{I}\,\delta^{(n)}(\vec{x} - \vec{y})\,,\\
    &\{\tilde{\phi}^{(1S)AB}(t\,,\vec{x})\,,\phi^{(1)\mu}(t\,,\vec{y})\} = 0\,.
\end{split}
\label{PB btw anti-symmetric / symmetric tilde phi^AB, tilde phi_A, and phi_AB-mu}
\end{equation}
Eq.~(\ref{PB btw tilde phi_AB and hat phi_AB}) and Eq.~(\ref{PB btw tilde phi^AB, tilde phi_A, and phi_AB-mu}) indicate that the primary constraint densities, $\phi^{(1)\mu} \approx 0$ given as the second formula in Eq.~(\ref{PC wrt eta}) and $\tilde{\phi}^{(1)}{}_{A} \approx 0$ given in Eq.~(\ref{modified PC wrt eta 2}) can be classified as first-class. To reveal this point, we need to perform the Dirac-Bergmann analysis~\cite{Dirac:1950pj,Dirac:1958sq,Bergmann:1949zz,BergmannBrunings1949,Bergmann1950,Anderson:1951ta} and investigate the PB-algebra among not only the primary constraint densities but also the secondary or higher-order constraint densities, if it exists. Notice also that the second and third PB-algebras in Eq.~(\ref{PB btw tilde phi_AB and hat phi_AB}) and the second PB-algebra in Eq.~(\ref{PB btw tilde phi^AB, tilde phi_A, and phi_AB-mu}), or equivalently, the second and third PB-algebra in Eq.~(\ref{PB btw anti-symmetric / symmetric tilde phi^AB, tilde phi_A, and phi_AB-mu}), vanish as the strong equality. This means that spacetime symmetries and internal-space symmetries can be formulated independently. To verify this statement, we need to perform the Dirac-Bergmann analysis to reveal the existence of secondary or higher-order constraint densities and calculate all PB-algebras among the constraint densities of the theory. 

Finally, let us consider the relation of the PB-algebras derived in this subsection to the Lie algebra of the affine group $A(n + 1\,;\,\mathbb{R}) = T(n + 1\,;\,\mathbb{R})\,\rtimes\,GL(n + 1\,;\,\mathbb{R})\,$, which satisfies
\begin{equation}
\begin{split}
    &[E^{IJ}\,,E^{KL}] = \eta^{IL}\,E^{KJ} - \eta^{KJ}\,E^{IL}\,,\\
    &[E^{IJ}\,,P_{K}] = \delta^{I}{}_{K}\,\eta^{JL}\,P_{L}\,,\\
    &[P_{I}\,,P_{J}] = 0\,,
\end{split}
\label{Affine Lie algebra}
\end{equation}
where $E^{IJ}$ and $P_{I}$ are the generators of the real general linear group $GL(n + 1\,;\,\mathbb{R})$ and of the translation $T(n + 1\,;\,\mathbb{R})\,$, respectively. Remark that $GL(n + 1\,;\,\mathbb{R})$ corresponds to the Lie group $G$ in the formulation introduced in Sec.~\ref{02:01}. The first algebra above exactly coincides with that of $\tilde{\phi}^{(1)AB}{}_{} \approx 0$ in Eq.~(\ref{PB btw tilde phi_AB in Formalism 1}) excepting the first term therein. Therefore, the internal STEGR in Formalism 1 can have internal symmetries generated only by the primary constraint densities given as the second formula in Eq.~(\ref{PC wrt eta}), Eq.~(\ref{modified PC wrt eta 2}), or higher-order constraint densities that are unveiled by performing the Dirac-Bergmann analysis. In particular, we would expect the appearance of a set of secondary constraint densities that restricts spacetime symmetries to such as the so-called hypersurface deformation algebra, which provides the diffeomorphism symmetry~\cite{Dirac:1958sc}. 

\section{\label{04}Internal STEGR in Formalism 2}
\subsection{\label{04:01}Canonical momenta and Primary constraints}
The Lagrangian density of the internal STEGR in Formalism 2 was introduced as Eq.~(\ref{STEGR Lagrangian in terms of e and eta in Formalism 2}) in Sec.~\ref{02:02}. The configuration space $\mathcal{Q}_{2}$ is coordinated by the two set of variables: $e_{A}{}^{\mu}$ and $\eta_{AB}\,$. In our notation, $T\mathcal{Q}_{2} = \left<e_{A}{}^{\mu}\,,\eta_{AB}\right>\,$. Thus, the velocity-phase space is provided by the tangent bundle of $\mathcal{Q}_{2}$, {\it i.e.,} $T\mathcal{Q}_{2} = \left<e_{A}{}^{\mu}\,,\eta_{AB}\,;\,\dot{e}_{A}{}^{\mu}\,,\dot{\eta}_{AB}\right>\,$. The phase-space is thus given by the dual-bundle of $T\mathcal{Q}_{2}$, {\it i.e.,} $T^{*}\mathcal{Q}_{2} = \left<e_{A}{}^{\mu}\,,\eta_{AB}\,;\,\pi^{A}{}_{\mu}\,,\pi^{AB}\right>\,$, where $\pi^{A}{}_{\mu}$ and $\pi^{AB}$ are canonical momenta with respect to each configuration variable, {\it i.e.,} $e_{A}{}^{\mu}$ and $\eta_{AB}\,$, given as follows:
\begin{equation}
    \pi^{A}{}_{\mu} = \frac{\delta \mathcal{L}_{\rm internal\,Formalism\,2}}{\delta \dot{e}_{A}{}^{\mu}} = 0\,
\label{CM wrt e in Formalism 2}
\end{equation}
and 
\begin{equation}
    \pi^{AB} = \frac{\delta \mathcal{L}_{\rm internal\,Formalism\,2}}{\delta \dot{\eta}_{AB}} = D^{ABEF}\,\dot{\eta}_{EF} + \frac{1}{2}\,e\,\sqrt{-\eta}\,e_{C}{}^{0}\,e_{D}{}^{i}\,\partial_{i}\eta_{EF}\,\tilde{\eta}^{CDABEF}\,,
\label{CM wrt eta in Formalism 2}
\end{equation}
respectively, where $D^{ABEF}$ and $\tilde{\eta}^{CDABEF}$ are defined by Eq.~(\ref{tilde_eta and D}) in Sec.~\ref{03:01}. Notice that Eq.~(\ref{CM wrt eta in Formalism 2}) is exactly the same as Eq.(\ref{CM wrt eta in Formalism 1}), but Eq.~(\ref{CM wrt e in Formalism 2}) is different from Eq.~(\ref{CM wrt e in Formalism 1}) in the existence of the additional term due to the vanishing-torsion property that is imposed in the Lagrangian density Eq.~(\ref{STEGR Lagrangian in terms of e and eta in Formalism 1}). Fundamental PB-algebra is introduced as follows:
\begin{equation}
\begin{split}
    &\{e_{A}{}^{\mu}(t\,,\vec{x})\,,\pi^{B}{}_{\nu}(t\,,\vec{y})\} = \delta_{A}{}^{B}\,\delta^{\mu}{}_{\nu}\,\delta^{(n)}(\vec{x} - \vec{y})\,,\\
    &\{\eta_{AB}(t\,,\vec{x})\,,\pi^{CD}(t\,,\vec{y})\} = \delta^{C}{}_{(A}\,\delta^{D}{}_{B)}\,\delta^{(n)}(\vec{x} - \vec{y})\,.
\end{split}
\label{Fundamental PB-algebra in Formalism 2}
\end{equation}
This PB-algebra is also the same as that in Formalism 1 investigated in Sec.~\ref{03:01} excepting the PB-algebra with respect to the additional term.  

Primary constraint density with respect to the frame field components $e_{A}{}^{\mu}$ in Formalism 2 are given as follows:
\begin{equation}
    \phi^{(1)A}{}_{\mu} := \pi^{A}{}_{\mu} \approx 0\,,
\label{PC wrt e}
\end{equation}
it is the same as that of Formalism 1 excepting the absence of the second term in Eq.~(\ref{PC wrt tau and e}). The total number of the components of the primary constraint density coincides with that of the upper bound: $(n + 1)^{2}$. To investigate internal symmetries in Sec.~\ref{04:02}, the above formula should be pulled back to the internal-space by the frame field, and the formula coincides with that of Formalism 1 given in Eq.~(\ref{modified PC wrt e in Formalism 1}), {\it i.e.,} $\tilde{\phi}^{(1)AB} := -\,\phi^{(1)A}{}_{\rho}\,\eta^{BI}\,e_{I}{}^{\rho} \approx 0\,$, excepting the absence of the additional term in that formula. We decompose it into the anti-symmetric and symmetric part in the same manner as the case of Formalism 1: Eq.~(\ref{modified PC1 wrt e in Formalism 1}), {\it i.e.,} $\phi^{(1aS)AB} := 2\,\tilde{\phi}^{(1)[AB]}\,$, and Eq.~(\ref{modified PC2 wrt e in Formalism 1}), {\it i.e.,} $\phi^{(1S)AB} := 2\,\tilde{\phi}^{(1)(AB)}\,$. In addition to the primary constraint density above, this sector has the primary constraint density given by either $\tilde{\phi}^{(1)A}{}_{BC} \approx 0$ in Eq.~(\ref{specific PC in Formalism 2 in internal-space}) or $\phi^{(1)\mu}{}_{BC} \approx 0$ in Eq.~(\ref{specific PC in Formalism 2 only in e}), which are introduced in Sec.~\ref{02:02}. It would be appropriate to choose the latter one since the former one contains the inverse frame field components: $\theta^{A}{}_{\mu}\,$. In our formalism, all quantities should be composed only of the frame field components $e_{A}{}^{\mu}\,$, the internal-space metric $\eta_{AB}\,$, and their canonical momenta $\pi^{A}{}_{\mu}$, $\pi_{AB}\,$. In the same manner as the case of Formalism 1, combining the primary constraint density given in Eq.~(\ref{specific PC in Formalism 2 only in e}), Eq.~(\ref{CM wrt eta in Formalism 2}) provides primary constraint density with respect to the internal-space metric $\eta_{AB}$ as follows:
\begin{equation}
    \phi^{(1)\mu} := \phi^{(1)\mu}{}_{IJ}\,\pi^{IJ} = -2\,e_{[I|}{}^{\rho}\partial_{\rho}e_{|J]}{}^{\mu}\,\pi^{IJ} \approx 0\,.
\label{PC wrt eta for spacetime in Formalism 2}
\end{equation}
The total number of the components of the primary constraint density is $(n + 1)$, and this number is less than that of the upper bound: $(n + 1)(n + 2)/2$. This constraint density is completely different from that of Formalism 1 given in the second formula of Eq.~(\ref{PC wrt eta}). Remark that Formalism 2 does not have the primary constraint density corresponding to $\tilde{\phi}^{(1)}{}_{A} \approx 0$ given as Eq.(\ref{modified PC wrt eta 2}). This makes sense since in Formalism 2 the imposition of the primary constraint density $\phi^{(1)\mu}{}_{IJ} \approx 0$ given in Eq.~(\ref{specific PC in Formalism 2 only in e}) make the torsion {\it a priori} vanishing.\footnote{
Remark that a primary constraint density holds as the weak equality in Hamiltonian formulation but as the strong equality in Lagrange formulation, although secondary or higher-order constraint densities do not have this property. 
}
Namely, the absence of the primary constraint density generating the translation symmetry indicates that in Formalism 2 the torsion does {\it a priori} not exist. In this point, we will briefly discuss in Sec.~\ref{05}, or more in detail, see Ref.~\cite{Tomonari:2023ars}. Therefore, it would be enough to investigate possible internal symmetries in $\tilde{\phi}^{(1)AB} \approx 0\,$, or equivalently, its anti-symmetric part $\phi^{(1aS)AB} \approx 0$ and symmetric part $\phi^{(1S)AB} \approx 0\,$, and $\phi^{(1)\mu} \approx 0\,$. To scrutinize this, we calculate the PB-algebra among these primary constraint densities in the next subsection. 

\subsection{\label{04:02}PB-algebras and Possible Symmetries}
We calculate PB-algebra of the primary constraint densities $\tilde{\phi}^{(1)AB} \approx 0\,$, $\tilde{\phi}^{(1)}{}_{A} \approx 0\,$, and $\phi^{(1)\mu} \approx 0\,$ to investigate possible internal symmetries in Formalism 2. In the first step, we calculate the PB-algebra in $\pi^{A}{}_{\mu}$-sector. Then, we reveal the PB-algebra in $(\pi^{(1)A}{}_{\mu}\,,\pi^{AB})$-sector. Finally, we consider the PB-algebra of $(\pi^{A}{}_{\mu}\,,\pi^{AB}\,;\,\phi^{(1)\mu}{}_{IJ})$-sector (the entire sector). 

The PB-algebra of $\pi^{A}{}_{\mu}$-sector is calculated as follows (See Appendix~\ref{App:02} for the derivation):
\begin{equation}
    \{\tilde{\phi}^{(1)AB}(t\,,\vec{x})\,,\tilde{\phi}^{(1)CD}(t\,,\vec{x})\} = \eta^{AD}\,\tilde{\phi}^{(1)CB} - \eta^{CB}\,\tilde{\phi}^{(1)AD}\,.
\label{PB btw tilde phi_AB in Formalism 2}
\end{equation}
Namely, the first term in Eq.~(\ref{PB btw tilde phi_AB in Formalism 1}) is absent in Formalism 2. Splitting this PB-algebra into the anti-symmetric and the symmetric part of $\tilde{\phi}^{(1)AB} \approx 0\,$, we obtain the following PB-algebras:
\begin{equation}
\begin{split}
    &\{\tilde{\phi}^{(1aS)AB}(t\,,\vec{x})\,,\tilde{\phi}^{(1aS)CD}(t\,,\vec{y})\} \\
    &\quad = \quad \left[\eta^{BD}\,\tilde{\phi}^{(1aS)AC} + \eta^{AC}\,\tilde{\phi}^{(1aS)BD} - \eta^{BC}\,\tilde{\phi}^{(1aS)AD} - \eta^{AD}\,\tilde{\phi}^{(1aS)BC}\right]\,\delta^{(n)}(\vec{x} - \vec{y})\,
\end{split}
\label{PB btw tilde phi_AB in anti-symmetric in Formalism 2}
\end{equation}
\begin{equation}
\begin{split}
    &\{\tilde{\phi}^{(1aS)AB}(t\,,\vec{x})\,,\tilde{\phi}^{(1S)CD}(t\,,\vec{y})\} \,\\
    &\quad = \quad \left[\eta^{BD}\,\tilde{\phi}^{(1S)AC} - \eta^{AC}\,\tilde{\phi}^{(1S)BD} +\,\eta^{BC}\,\tilde{\phi}^{(1S)AD} - \eta^{AD}\,\tilde{\phi}^{(1S)BC}\right]\,\delta^{(n)}(\vec{x} - \vec{y})\,
\end{split}
\label{PB btw tilde phi_AB in anti-symmetric and symmetric in Formalism 2}
\end{equation}
and
\begin{equation}
\begin{split}
    &\{\tilde{\phi}^{(1S)AB}(t\,,\vec{x})\,,\tilde{\phi}^{(1S)CD}(t\,,\vec{y})\} \,\\
    &\quad = \quad \left[\eta^{BD}\,\tilde{\phi}^{(1aS)AC} + \eta^{AC}\,\tilde{\phi}^{(1aS)BD} + \eta^{BC}\,\tilde{\phi}^{(1aS)AD} + \eta^{AD}\,\tilde{\phi}^{(1aS)BC}\right]\,\delta^{(n)}(\vec{x} - \vec{y})\,.
\end{split}
\label{PB btw tilde phi_AB in symmetric in Formalism 2}
\end{equation}
Differing from Formalism 1, the primary constraint density $\tilde{\phi}^{(1)AB} \approx 0$ can be classified as first-class. To verify this, we should investigate the PB-algebra of $\tilde{\phi}^{(1)AB} \approx 0$ among other primary constraint densities, {\it i.e.,} $\tilde{\phi}^{(1)}{}_{A} \approx 0\,$, and $\phi^{(1)\mu} \approx 0\,$ and, secondary and higher-order constraint density, if it exists. In particular, the latter case needs to perform the Dirac-Bergmann analysis. If the primary constraint density $\tilde{\phi}^{(1)AB} \approx 0$ are classified as first-class, then the PB-algebra given by Eq.~(\ref{PB btw tilde phi_AB in Formalism 2}) coincides with the first algebra of the affine Lie algebra of Eq.~(\ref{Affine Lie algebra}). In particular, Eq.~(\ref{PB btw tilde phi_AB in anti-symmetric in Formalism 2}) is nothing but the Lorentz algebra. Therefore, Formalism 2 has a possibility to hold the internal symmetry, which is the same as that of GR, as desired. 

The PB-algebra of $(\pi^{A}{}_{\mu}\,,\pi^{AB}\,;\,\phi^{(1)\mu}{}_{IJ})$-sector (the entire sector) is calculated as follows (See Appendix~\ref{App:02} for the derivation):
\begin{equation}
\begin{split}
    &\{\phi^{(1)\mu}(t\,,\vec{x})\,,\phi^{(1)\nu}(t\,,\vec{y})\} = 0\,,\\
    &\{\phi^{(1)\mu}(t\,,\vec{x})\,,\tilde{\phi}^{(1)AB}(t\,,\vec{y})\} = 0\,.
\end{split}
\label{PB btw phi mu and tilde phi in Formalism 2}
\end{equation}
Equivalently, the second PB-algebra above is split into the anti-symmetric and the symmetric part as follows:
\begin{equation}
\begin{split}
    &\{\phi^{(1)\mu}(t\,,\vec{x})\,,\tilde{\phi}^{(1aS)AB}(t\,,\vec{y})\} = 0\,,\\
    &\{\phi^{(1)\mu}(t\,,\vec{x})\,,\tilde{\phi}^{(1S)AB}(t\,,\vec{y})\} = 0\,.
\end{split}
\label{PB btw phi mu and tilde phi in anti-symmetric and symmetric in Formalism 2}
\end{equation}
These PB-algebras hold as the strong equality. This means that spacetime symmetries and internal-space symmetries can be established independently. We would expect that the Dirac-Bergmann analysis provides spacetime symmetry such as the diffeomorphism symmetry in the secondary constraint densities. 

\section{\label{05}Conclusions}
In this work, we revisited STEGR based on the gauge approach to gravity and verified that the non-metricity automatically vanishes in the use of the ordinary Minkowskian internal-space metric. To overcome this issue, introducing a generic internal-space metric that is variable in the internal-space, we established three Formalisms. In this approach, since the torsion does not vanish automatically, it was necessary to impose some conditions for the vanishing. In Formalism 1, using auxiliary variables, the vanishing condition was implemented. In Formalism 2, the condition was taken into account as a primary constraint of the theory. Formalism 3 was somewhat special: we introduced a specific decomposition of the (co-)frame field components by using the so-called St\"{u}ckelberg fields to realize {\it a priori} vanishing the torsion. In this formalism, we found a weird structure like a bi-metric structure. In this paper, we focused on investigating the internal-space symmetry of the internal STEGR in Formalism 1 and Formalism 2. In the internal STEGR in Formalism 1, on one hand, we found that the translation symmetry given by $T(n + 1\,;\,\mathbb{R})$ can be held and the symmetries provided by $GL(n+1\,;\,\mathbb{R})\,$, in which the local Lorentz symmetry contains, are broken. On the other hand, in the internal STEGR in Formalism 2, we found that the translation symmetry is {\it a priori} absent and all the symmetries provided by $GL(n+1\,;\,\mathbb{R})$ can be held. 

In the previous work, the author proposed a unified description of metric-affine geometries using the M\"{o}bius representation, in which all geometric quantities such as curvature, torsion, and non-metricity are formulated on the same ground using the M\"{o}bius representation~\cite{Tomonari:2023ars}. In this unified description, a restriction to the translation symmetry, or in a stronger statement, a case that the theory does {\it a priori} not have any generator of the translation symmetry, implies that the torsion may vanish. The internal STEGR in Formalism 2 would, on one hand, be a suitable theory in this perspective since the theory does {\it a priori} not have the translation symmetry while possibly holding at least the local Lorentz symmetry. On the other hand, the internal STEGR in Formalism 1 can satisfy the translation symmetry without any restriction, and this might be inconsistent with the unified description unless some extra gauge condition on the Cartan connection is imposed. Another possibility to reconcile this situation is that the Dirac-Bergmann procedure provides a set of secondary or higher-order constraint densities such that the translation symmetry is broken. In addition, in Formalism 1 and Formlism 2, we do not find the generator of dilation and shear, or equivalently, non-metricity in the presented work. To be consistent with the previous work~\cite{Tomonari:2023ars}, the generator, which would be the trace and trace-free part of the symmetric generator of some symmetries in non-metricity, respectively, should be discovered. Finally, we should not overlook a novel fact addressed in Ref.~\cite{Obukhov:2024evf} that a violation of Lorentz symmetry has something to do with the emergence of nontrivial non-metricity and also its detection in physical observations due to some change in the notion of light cone.

For future perspectives, to determine the symmetries in the internal STEGR in Formalism 1 and Formalism 2, we have to perform the Dirac-Bergmann analysis to unveil all secondary or higher-order constraint densities and calculate all PB-algebras among them. We should also find the spacetime symmetries in Formalism 1 and Formalism 2, which would appear in secondary constraint densities, such as the hypersurface deformation algebra~\cite{Dirac:1958sc}. In addition, it is mandatory to clarify the relation between the origin of non-metricity and some symmetry breaking in detail, as already addressed in Ref.~\cite{Obukhov:2024evf}. It would give another perspective on that origin from the viewpoint of the unified description proposed in Ref.~\cite{Tomonari:2023ars} and vice versa. Based on these investigations, we would formulate a consistent theory of internal STEGR with the unified description proposed in Ref.~\cite{Tomonari:2023ars}, and the theory would invite us to a new stage of understanding the theories of STEGR and, ultimately, unveiling the nature of gravity. 

\begin{acknowledgments}
KT would like to thank Sebastian Bahamonde and Taishi Katsuragawa for giving beneficial comments for this work and the cosmology theory group in Institute of Science Tokyo for supporting my work, in particular, professor Teruaki Suyama.
\end{acknowledgments}

\section*{Declarations}
\section*{Data availability}
Data sharing not applicable to this article as no datasets were generated or analysed during the current study.
\section*{Conflicts of interests}
The author have no competing interests to declare that are relevant to the content of this article.

\appendix
\section{\label{App:01}Complementary PB-algebra for internal STEGR in Formalism 1}
The calculation of the PB-algebra of $\pi^{A}{}_{\mu}$- and $(\pi^{A}{}_{\mu}\,,\pi_{AB}{}^{\mu})$-sector need the following complementary PB-algebras:
\begin{equation}
\begin{split}
    &\{\phi^{(1)A}_{\,\,\,\,\,\,\,\,\,\,\,\,\mu}(t\,,\vec{x})\,,\phi^{(1)B}_{\,\,\,\,\,\,\,\,\,\,\,\,\mu}(t\,,\vec{y})\} = -\,4\,\tau^{AB}{}_{(\mu}\delta_{\nu)}{}^{0}\,\delta^{(n)}(\vec{x} - \vec{y})\,,\\
    &\{\phi^{(1)\,\,\,\,\,\,\,\,\,\,\mu}_{\,\,\,\,\,\,\,AB}(t\,,\vec{x})\,,\phi^{(1)\,\,\,\,\,\,\,\,\,\,\mu}_{\,\,\,\,\,\,\,CD}(t\,,\vec{y})\} = 0\,,\\
    &\{\phi^{(1)A}_{\,\,\,\,\,\,\,\,\,\,\,\,\mu}(t\,,\vec{x})\,,\phi^{(1)\,\,\,\,\,\,\,\,\,\,\nu}_{\,\,\,\,\,\,\,BC}(t\,,\vec{y})\} = 2\,e_{[B}{}^{0}\,\delta^{A}{}_{C]}\,\delta_{\mu}{}^{\nu}\,\delta^{(n)}(\vec{x} - \vec{y})\,,\\
    &\{\tilde{\phi}^{(1)AB}(t\,,\vec{x})\,,\phi^{(1)\,\,\,\,\,\,\,\,\,\,\mu}_{\,\,\,\,\,\,\,CD}(t\,,\vec{y})\} = 2\,\eta^{BI}\,e_{I}{}^{\mu}\,e_{[C}{}^{0}\,\delta_{D]}{}^{A}\,\delta^{(n)}(\vec{x} - \vec{y})\,,\\
    &\{\phi^{(1)\,\,\,\,\,\,\,\,\,\,\mu}_{\,\,\,\,\,\,\,AB}(t\,,\vec{x})\,,\phi^{(1)\,\,\,\,\,\,\,\,\,\,\nu}_{\,\,\,\,\,\,\,CD}(t\,,\vec{y})\} = 0\,.
\end{split}
\label{comp PB btw pi^A_mu and pi_AB^mu}
\end{equation}
Using Eq.~(\ref{PB btw tilde phi_AB in Formalism 1}) and the fourth formula in Eq.~(\ref{comp PB btw pi^A_mu and pi_AB^mu}), the following PB-algebras are derived
\begin{equation}
\begin{split}
    &\{\phi^{(1aS)AB}(t\,,\vec{x})\,,\phi^{(1)\,\,\,\,\,\,\,\,\,\,\mu}_{\,\,\,\,\,\,\,CD}(t\,,\vec{y})\} = 4\,e_{I}{}^{\mu}\,e_{[C}{}^{0}\,\eta^{I[A}\,\delta_{D]}{}^{B]}\,\delta^{(n)}(\vec{x} - \vec{y})\,,\\
    &\{\phi^{(1S)AB}(t\,,\vec{x})\,,\phi^{(1)\,\,\,\,\,\,\,\,\,\,\mu}_{\,\,\,\,\,\,\,CD}(t\,,\vec{y})\} = -\,4\,e_{I}{}^{\mu}\,e_{[C}{}^{0}\,\eta^{I(A}\,\delta_{D]}{}^{B)}\,\delta^{(n)}(\vec{x} - \vec{y})\,.
\end{split}
\label{}
\end{equation}

The calculation of the PB-algebra of $(\pi^{A}{}_{\mu}\,,\pi_{AB}{}^{\mu}\,,\pi^{AB})$-sector (the entire sector) needs the following complementary PB-algebras:
\begin{equation}
\begin{split}
    &\{C_{AB\mu}(t\,,\vec{x})\,,\pi^{CD}(t\,,\vec{y})\} = 2\,\eta_{[A|I|}\,\tau^{I(C}{}_{\mu}\,\delta^{D)}{}_{|B]}\,\delta^{(n)}(\vec{x} - \vec{y})\,\\
    &\{C_{AB\mu}(t\,,\vec{x})\,,C_{AB\mu}(t\,,\vec{y})\} = 0\,.
\end{split}
\label{}
\end{equation}
Using these PB-algebras, the PB-algebra of $\phi^{(1)}{}_{\mu}$ are calculated as follows:
\begin{equation}
\begin{split}
    &\{\phi^{(1)}{}_{\mu}(t\,,\vec{x})\,,\phi^{(1)}{}_{\nu}(t\,,\vec{y})\} = \\
    &\quad\left[-\,2\,C_{IJ\mu}\,\pi^{KL}\,\eta_{[K|A|}\,\tau^{A(I}{}_{\nu}\,\delta^{J)}{}_{|L]} + \,2\,C_{IJ\nu}\,\pi^{KL}\,\eta_{[K|A|}\,\tau^{A(I}{}_{\mu}\,\delta^{J)}{}_{|L]}\right]\,\delta^{(n)}(\vec{x} - \vec{y}) = 0\,.
\end{split}
\label{}
\end{equation}
The PB-algebra either of $e_{A}{}^{\mu}$ or $\eta_{AB}$ and $\phi^{(1)}{}_{\mu}$ are calculated as follows:
\begin{equation}
\begin{split}
    &\{e_{A}{}^{\mu}(t\,,\vec{x})\,,\phi^{(1)}{}_{\nu}(t\,,\vec{y})\} = 0\,,\\
    &\{\eta_{AB}(t\,,\vec{x})\,,\phi^{(1)}{}_{\mu}(t\,,\vec{y})\} = 0\,.\\
\end{split}
\label{}
\end{equation}

The calculation of the PB-algebra of $(\pi^{A}{}_{\mu}\,,\pi_{AB}{}^{\mu}\,,\pi^{AB})$-sector (the entire sector) needs the following complementary PB-algebras:
\begin{equation}
\begin{split}
    &\{\phi^{(1)}{}_{\nu}(t\,,\vec{x})\,,\phi^{(1)A}_{\,\,\,\,\,\,\,\,\,\,\,\,\mu}(t\,,\vec{y})\} = 0\,,\\
    &\{\phi^{(1)}{}_{\mu}(t\,,\vec{x})\,,\phi^{(1)\,\,\,\,\,\,\,\,\,\,\nu}_{\,\,\,\,\,\,\,AB}(t\,,\vec{y})\} = \pi^{IJ}\,\eta_{IK}\,\eta_{JL}\,\delta^{K}{}_{[A}\,\delta^{L}{}_{B]}\,\delta_{\mu}{}^{\nu}\,\delta^{(n)}(\vec{x} - \vec{y}) = 0\,,\\
    &\{\tilde{\phi}^{(1)}{}_{A}(t\,,\vec{x})\,,\phi^{(1)\,\,\,\,\,\,\,\,\,\,\mu}_{\,\,\,\,\,\,\,BC}(t\,,\vec{y})\} = 0\,,\\
    &\{\phi^{(1)\mu}(t\,,\vec{x})\,,\phi^{(1)\,\,\,\,\,\,\,\,\,\,\nu}_{\,\,\,\,\,\,\,BC}(t\,,\vec{y})\} = 0\,.
\end{split}
\label{}
\end{equation}
Here, we used the following PB-algebras:
\begin{equation}
\begin{split}
    &\{\pi^{AB}(t\,,\vec{x})\,,\phi^{(1)C}_{\,\,\,\,\,\,\,\,\,\,\,\,\mu}(t\,,\vec{y})\} = 0\,,\\
    &\{\pi^{AB}(t\,,\vec{x})\,,\phi^{(1)\,\,\,\,\,\,\,\,\,\,\mu}_{\,\,\,\,\,\,\,CD}(t\,,\vec{y})\} = 0\,
\end{split}
\label{}
\end{equation}
and
\begin{equation}
\begin{split}
    &\{C_{AB\mu}(t\,,\vec{x})\,,\phi^{(1)A}_{\,\,\,\,\,\,\,\,\,\,\,\,\nu}(t\,,\vec{y})\} = 0\,,\\
    &\{C_{AB\mu}(t\,,\vec{x})\,,\phi^{(1)\,\,\,\,\,\,\,\,\,\,\nu}_{\,\,\,\,\,\,\,CD}(t\,,\vec{y})\} = \eta_{AI}\,\eta_{BJ}\,\delta^{I}{}_{[C}\,\delta^{J}{}_{D]}\,\delta_{\mu}{}^{\nu}\,\delta^{(n)}(\vec{x} - \vec{y})\,.
\end{split}
\label{}
\end{equation}

\section{\label{App:02}Complementary PB-algebra for internal STEGR in Formalism 2}
The calculation of the PB-algebra of $\pi^{A}{}_{\mu}$- and $(\pi^{A}{}_{\mu}\,,\pi^{AB}\,;\,\phi^{(1)\mu}{}_{AB})$-sector (the entire sector) needs the following complementary PB-algebras:
\begin{equation}
\begin{split}
    &\{\phi^{(1)A}{}_{\mu}(t\,,\vec{x})\,,\phi^{(1)B}{}_{\nu}(t\,,\vec{y})\} = 0\,,\\
    &\{\phi^{(1)\mu}{}_{AB}(t\,,\vec{x})\,,\phi^{(1)\nu}{}_{CD}(t\,,\vec{x})\} = 0\,,\\
    &\{\phi^{(1)A}{}_{\mu}(t\,,\vec{x})\,,\phi^{(1)\nu}{}_{BC}(t\,,\vec{x})\} = 2\,\delta^{A}{}_{[B|}\,\partial_{\mu}e_{|C]}{}^{\nu}\,\delta^{(n)}(\vec{x} - \vec{y}) + 2\,\delta^{\mu}{}_{\nu}\,\delta^{A}{}_{[B}e_{C]}{}^{\rho}\,\partial_{\rho}\delta^{(n)}(\vec{x} - \vec{y})\,.
\end{split}
\label{}
\end{equation}

The calculation of the PB-algebra of $(\pi^{A}{}_{\mu}\,,\pi^{AB}\,;\,\phi^{(1)\mu}{}_{AB})$-sector (the entire sector) needs the following complementary PB-algebras:
\begin{equation}
\begin{split}
    &\{\phi^{(1)\mu}{}_{AB}(t\,,\vec{x})\,,\pi^{CD}(t\,,\vec{y})\} = 0\,,\\
    &\{\phi^{(1)\mu}(t\,,\vec{x})\,,\eta_{AB}(t\,,\vec{y})\} = 0\,,\\
    &\{\phi^{(1)\mu}(t\,,\vec{x})\,,\phi^{(1)A}{}_{\nu}(t\,,\vec{y})\} = 0\,,\\
    &\{\phi^{(1)\mu}(t\,,\vec{x})\,,\phi^{(1)\nu}{}_{AB}(t\,,\vec{y})\} = 0\,.
\end{split}
\label{}
\end{equation}

\section*{References}
\bibliography{bibliography.bib}
\bibliographystyle{unsrt}
\end{document}